\documentclass[fleqn,usenatbib]{mnras}
\usepackage{newtxtext,newtxmath}
\usepackage[T1]{fontenc}
\DeclareRobustCommand{\VAN}[3]{#2}
\let\VANthebibliography\thebibliography
\def\thebibliography{\DeclareRobustCommand{\VAN}[3]{##3}\VANthebibliography}

\usepackage{graphicx}	
\usepackage{amsmath}	
	
\usepackage{color}
\usepackage{subcaption}
\usepackage{wrapfig}
\usepackage{gensymb}

\title[Accreted stellar streams]{The structure of accreted stellar streams}

\author[Qian, Arshad, \& Bovy]{
Yansong Qian\thanks{yansong.qian@mail.utoronto.ca},
Yumna Arshad,
and Jo Bovy
\\
\phantom{$^1$}David A. Dunlap Department of Astronomy and Astrophysics, University of Toronto, 50 St. George Street, Toronto, Ontario, M5S 3H4, Canada
}

\date{}

\pubyear{2021}

\begin{document}
\label{firstpage}
\pagerange{\pageref{firstpage}--\pageref{lastpage}}
\maketitle

\begin{abstract}
Many of the Milky Way's globular clusters are likely accreted from satellite galaxies that have long since merged with the Milky Way. When these globular clusters are susceptible to tidal disruption, this process likely starts already inside the parent satellite leading to an early stellar stream within the satellite. When the parent satellite merges with the Milky Way, the globular cluster and its pre-merger stellar stream are accreted in a somewhat chaotic process. Here, we investigate the properties of the accreted stream after the merger as we would see it today using a suite of simulations of accretion events. We find that the accretion process leads to a wide range of behaviors, but generally scatters the accreted stream over a wide, two-dimensional area of the sky. The behavior ranges from a set of a few or more well-defined ''sub-streams'' extending out from the post-merger thin stream by tens of degrees, to more widely dispersed debris over much of the sky, depending on how close to the center of the Milky Way the merger happened. Using mock \emph{Gaia}-like observations of the simulated streams, we demonstrate that an accreted-stream component can explain the off-track features observed in the GD-1 stream. Sub-streams can appear like thin tidal streams themselves that are seemingly unassociated with the post-merger stream, raising the possibility that some of the progenitor-less streams observed in the Milky Way are part of a single or a few accreted streams created in an ancient merger event.
\end{abstract}


\section{Introduction}

Globular clusters lose stars when they orbit galaxies due to the tidal effect. The stripped stars are ejected with small initial velocities with respect to the globular-cluster progenitor and form a one-dimensional stream \citep[e.g.,][]{Johnston1998,Eyre2011,Bovy2014}. Many thin tidal streams have been found observationally in the last decades \citep[e.g.,][]{Odenkirchen2001,Grillmair2006,ibata2020}. However, many globular clusters around the Milky Way likely originated in satellite galaxies that merged with the Milky Way \citep[e.g.,][]{Leaman2013, Forbes2018, Kruijssen2019} and these globular clusters orbit their original hosts until these dwarf galaxies are tidally disrupted by the Milky Way potential. The accreted streams formed by this system are expected to display more complex behavior than the simple thin streams that form in a smooth, static mass distribution, because they evolve in the complex, time-dependent gravitational potential of the dwarf galaxy and the Milky Way \citep{Malhan2019,Carlberg2020}, which could allow us to constrain the inner density profile of the dwarf galaxy \citep[e.g.,][]{Malhan2021}.

Thin tidal streams are of great interest because they can be used to determine the large-scale mass distribution of the Milky Way halo \citep{Johnston1999} including its shape \citep[e.g.,][]{Bovy2016}. Furthermore, thin tidal streams are so sensitive to perturbations that they can be used to build up a detailed picture of small-scale inhomogeneities in the Galactic potential, including those resulting from starless dark-matter subhalos \citep{Ibata2002,Johnston2002,Carlberg2012,Amorisco2016,Erkal2016,Bovy2017,Pearson2017,Banik2019,Banik2021}. An accreted stream carries information about the merger of its parent galaxy and the Milky Way in the past, which is an important process in the formation of galaxies. Finding and characterizing signatures of accreted streams in the Milky Way would therefore help improve our understanding of the merger history of our Galaxy and increase the fidelity with which we can use tidal streams to, e.g., test theories of the particle nature of dark matter \citep{Banik2019b}. This is complementary to other probes of the Milky Way's accretion history, such as using the orbits and elemental abundances of stellar halo stars to identify accreted components \citep{Helmi2018, Belokurov2018} or using globular clusters as tracers of accretion events \citep{Myeong2018, Kruijssen2020}.

While many works in the literature have investigated the structure of tidal streams formed in different smooth potentials and under the influence of different perturbers, few simulations have been performed of accreted streams formed around accreted globular clusters. \citet{Carlberg2020} conduct full $N$-body simulations of globular clusters started in dwarf galaxies that later merge with a Milky-Way-like host galaxy in a cosmological simulation. They find that the accreted component of the stream is dispersed after the merger as a chaotic thick stream with a width roughly that of the cluster inside the dwarf galaxy. Motivated by the discovery of a ``cocoon'' structure in the GD-1 stream, \citet{Malhan2019} simulate accreted streams in a similar way as \citet{Carlberg2020} and find that there are always diffuse stellar components in the formed mock streams. These components have a wide range of size and behavior, but their work does not involve a detailed characterisation of the types of accreted structure that can be formed.

In this paper, we use test-particle simulations to model the formation process of accreted streams starting from their evolution inside the parent satellite galaxies to several Gyrs after they are become bound to the Milky Way. The initial conditions of our simulation are based on the orbits of observed satellite galaxies of the Milky Way. We discuss our simulation methods in Section \ref{sec:method}. In Section \ref{sec:results}, we study the structure of the accreted streams formed in our simulations, determine the impact of variations in the merger process, and look at the action distribution of the accreted streams to better understand the origin of the observed structure. In Section \ref{sec:observations}, we create mock observations similar to those of the GD-1 stream to investigate whether the features seen in our simulations can explain the observed features in the GD-1 stream and we determine the impact of the accreted stream on interpretations of the density structure of thin streams. We present our conclusions in Section \ref{sec:conclusions}.

\section{Method}\label{sec:method}

To achieve our aim of exploring the diversity of structure around stellar streams originating from accreted globular clusters, we perform a suite of test-particle simulations of stream formation. Using test-particle simulations rather than full $N$-body simulations allows us to quickly explore different outcomes depending on the orbit of the progenitor globular cluster and its parent dwarf galaxy, the time at which the globular cluster gets stripped from its parent dwarf, and the initial location of the globular cluster within the dwarf. We can also easily employ realistic models for the Milky Way and dwarfs' gravitational potentials. As a proxy for possible parent-dwarf orbits, we use the orbits of the present-day satellite galaxies of the Milky Way. Integrating these backwards while taking into account the effect of dynamical friction, we arrive at plausible initial conditions for a globular cluster that accretes onto the Milky Way when its parent dwarf merges with it. From these initial conditions, we simulate the forward evolution of the stream--globular-cluster system in the combined Milky-Way--dwarf potential using a particle-spray technique and finally produce simulated accreted streams as they would appear today. All of these steps of our methodology are described in detail in the following subsections.

\subsection{Orbit of progenitor}

To create plausible accretion scenarios for the globular-cluster progenitors of accreted stellar streams, we place globular clusters in dwarf galaxies at different times along the past orbit of current satellite galaxies of the Milky Way. We stress that we only use the current satellites to obtain a plausible orbit of an accreted globular cluster's parent galaxy, not to simulate the accretion of globular clusters from present-day dwarf satellites. While the orbits of now-accreted dwarf galaxies were likely more radial and eccentric than those of the surviving dwarf galaxies that we use, we find no strong effect of the eccentricity of the parent dwarf on the accreted streams that we simulate. Therefore, our use of surviving dwarf galaxies does not create a strong bias in our results. 

To begin, take the present-day position and velocity of the dwarf galaxies within 420 kpc from \citet{Fritz2018}, except that we use the values from \citet{Helmi18} for the LMC, SMC, and Sgr (see \citealt{Garrow2020}). To integrate their orbits backward in time, we use the Python galactic dynamics package \texttt{galpy}\footnote{Available at \url{http://github.com/jobovy/galpy}~.} \citep{galpy} using the  \texttt{MWPotential2014} mass model from \citep{galpy}. We include the effect of dynamical friction using an implementation in \texttt{galpy} comparable to that of \citet{Petts2016}; here and below, we assign each dwarf galaxy a total mass of $10^{11}\,M_\odot$ and model its mass profile as a \citet{Hernquist1990} profile with a scale length of 33.2 kpc. We integrate each dwarf galaxy for 10 Gyr backwards in time.

Starting from their position and velocity 10 Gyr ago, we integrate the orbits forward in time for up to 15 Gyr and select satellite orbits and a 5 Gyr segment of them that satisfy the following conditions:

\begin{enumerate}
         \item The orbit is bound to the Milky Way and reaches $<40$ kpc from the center at the end of the integration period. We require this, because we assume that the satellite needs to be closer than 40 kpc in order to get tidally disrupted and place a globular cluster on an inner-halo orbit. The tidal disruption of the satellite galaxy brings the accreted globular cluster to the Milky Way. 
         
         \item The orbit has an eccentricity $\lesssim 0.8$. This helps with producing accreted globular-clusters that go on realistic stream-producing orbits around the Milky Way.
         
         \item For analysis purposes, especially for studying the action distribution of the stream, we prefer satellite galaxies with orbits in $x$-$y$ plane rather than in the $x$-$z$ or $y$-$z$ plane (where $z$ is the disk axis). This choice, done purely for visualization reasons, allows us to marginalize over $J_z$ and study the action distribution in 2D. This property is determined manually, so there is no hard criterion. 
     \end{enumerate}
     
In the end, this set of criteria leaves us with the past orbits of Fornax, Draco, Sextans, and Willman 1 and these are the past orbits that we use in the remainder of the paper.

With the satellite orbits in hand, we then go back to the beginning of the 5 Gyr segment selected above and initialize a globular cluster on a circular orbit at either 2 or 4 kpc from the center of the dwarf galaxy (similar to the prescription used by \citealt{Carlberg2018a}), modeled as a Hernquist profile with a mass of $10^{11}\,M_\odot$ and scale radius 33.2 kpc. We integrate the orbit of the globular cluster forward in time in the combined potential of the Milky Way and the dwarf galaxy. For the first 5 Gyr of integration, the mass of the dwarf galaxy is kept constant and the globular cluster remains bound to the dwarf galaxy as the latter's orbit decays under the effect of dynamical friction. After 5 Gyr, when the satellite is closer than 40 kpc from the center of the Milky Way, we smoothly drop the mass of the dwarf galaxy to zero using the smoothing prescription from \citet{dehnen2000} over a period of 0.5 Gyr. This simulates the tidal disruption of the dwarf galaxy. After this, the globular cluster orbits for another 4.5 Gyr in the Milky Way potential alone. Thus, the globular cluster orbits for a total of 10 Gyr, 5 Gyr bound to the dwarf galaxy, 0.5 Gyr during the dwarf's tidal disruption, and 4.5 Gyr in the Milky Way alone.

\begin{figure}
    \centering
    \includegraphics[width =\linewidth]{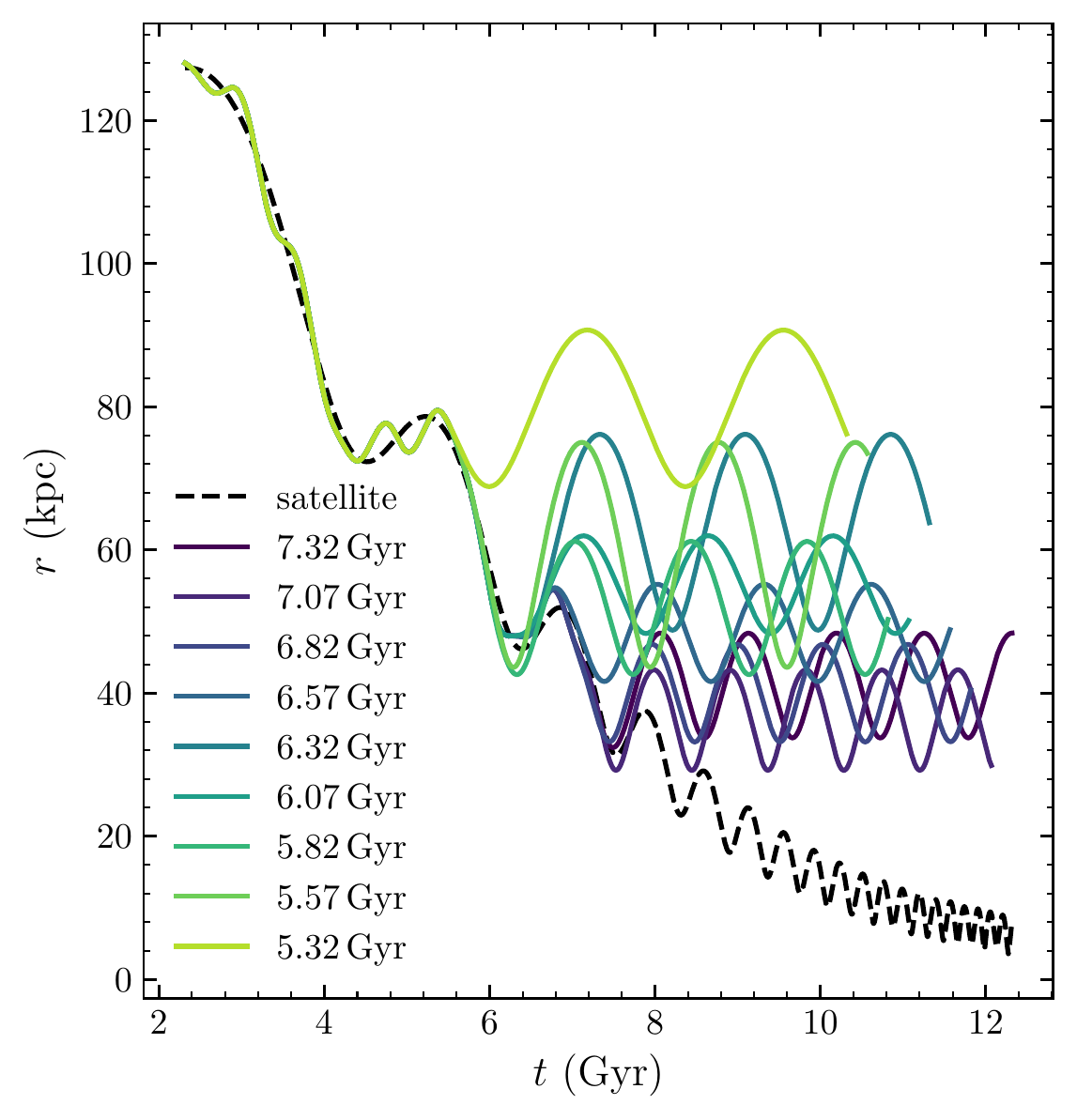}
    \caption{The orbit of globular clusters generated using the orbit of Fornax (dashed curve) with different merger times---indicated by different colors---at which the globular cluster is removed from its parent satellite. $r$ indicates the distance from the Milky Way center.}
    \label{fig:sc orbit}
\end{figure}

Example accreted globular-cluster orbits produced using the past orbit of Fornax are shown in Figure \ref{fig:sc orbit}. Different lines show different times at which we let the dwarf galaxy's mass go to zero, thus resulting in different orbits of the accreted globular cluster. Among these globular clusters, only the darkest purple one is unbound when the Fornax reaches 40 kpc, and the others are stripped before that because we start to lower the mass of Fornax earlier.

\subsection{Generating stellar streams}

Compared to the satellite dwarf galaxy, the tidal disruption of the globular cluster proceeds much more slowly. We assume that the globular cluster starts tidally disrupting within its parent dwarf galaxy and continues to lose stars after it transitions into orbiting within the Milky Way alone after the tidal disruption of its parent dwarf; this is plausible because cosmological zoom-in simulations demonstrate that accreted globular clusters typically settle on orbits with similar tidal fields as they experienced within their original host \citep{Carlberg2018a}. The second phase of tidal disruption produces a thin stellar stream, similar to the GD-1 or Pal 5 streams; the present-day structure of the part of the stream produced inside the parent dwarf is the object of this study.

To simulate stream formation, we use the particle-spray method from \citet{stream} as implemented in the \texttt{streamtools}\footnote{Available at \url{http://github.com/jobovy/streamtools}~.} package. This implementation was first used in \citet{Banik2019}, but we adapt it here for streams generated by clusters orbiting in different systems (dwarf galaxy and Milky Way) along their past orbits. The method from \citet{stream} consists of a prescription for the positional and velocity offset of a tidally-stripped star from its progenitor cluster using coordinates based on the orbital plane of the progenitor (the method was developed for spherical potentials where orbits have a fixed orbital plane). Depending on whether the velocity offset is in the direction of or against the direction of rotation determines whether a star goes into the leading or trailing stream; we generate both with equal numbers of stars. We adapt the prescription from \citet{stream} as follows: (i) we apply it in the instantaneous orbital plane calculated with respect to the center of the system that the progenitor is orbiting and (ii) we assume that the progenitor is orbiting solely around the dwarf's center before the start of the dwarf's tidal disruption and solely around the Milky Way after the end of the dwarf's tidal disruption. To avoid the ambiguity of what to do during tidal disruption, we assume that no stars are stripped from the globular cluster during this time. The instantaneous orbital plane is computed based on the relative angular momentum of the globular cluster with respect to its orbital center. Thus, before tidal distribution, it is computed relative to the moving dwarf's center and after relative to the static Milky Way potential. 

In the instantaneous orbital-plane frame, the generated stream stars are given positions and velocities according to \citep{stream}
\begin{equation}\begin{split}
r &= r_\mathit{sc} + k_r \, r_\mathrm{tidal} \\ 
\phi &= \phi_\mathit{sc} + k_\phi \, r_\mathrm{tidal} / r \\
v_r &= v_{r,\mathit{sc}} + k_{vr} \, v_{r,\mathit{sc}} \\
v_t &= v_{t,\mathit{sc}} + k_{v\phi} \, V_c(r_\mathit{sc}) r_\mathrm{tidal} / r  \\
z &= k_z \, r_\mathrm{tidal} / r  \\
v_z &= k_{vz} \, V_c(r_\mathit{sc}) r_\mathrm{tidal} / r\,, \label{eq streamvel}
\end{split}\end{equation}
where quantities with a `sc' subscript denote the properties of the progenitor globular cluster. We follow the specific model described in \citet{stream}, where $k_\phi = k_{vr}  = 0$ and the other $k_i$s are randomly sampled from Gaussian distributions $N(\mu,\sigma)$ with mean $\mu$ and standard deviation $\sigma$:
\begin{equation}\begin{split}
k_\phi &= k_{vr}= 0\\
k_r &\sim N(2,0.4)\\
k_{vt} &\sim N(0,0.5)\\
k_{z} &\sim N(0,0.5)\\
k_{vz} &\sim N(0.3,0.4)\,.
\label{eq streamvel k}
\end{split}\end{equation}
Further in Equation \eqref{eq streamvel}, $V_c$ is the circular velocity and $r_\mathit{sc}$ is the tidal radius of the globular cluster. These are again computed based either on the parent dwarf galaxy or the Milky Way depending on whether the stream star is generated before or after the tidal disruption of the dwarf. For the calculation of the tidal radius, we use a globular-cluster mass of $20,000\,M_\odot$. After a stream star is generated using this prescription, its orbit is integrated in the combined potential of the dwarf galaxy and the Milky Way, or just the Milky Way after the tidal disruption of the dwarf.

\begin{figure}
    \centering
    \includegraphics[width =\linewidth]{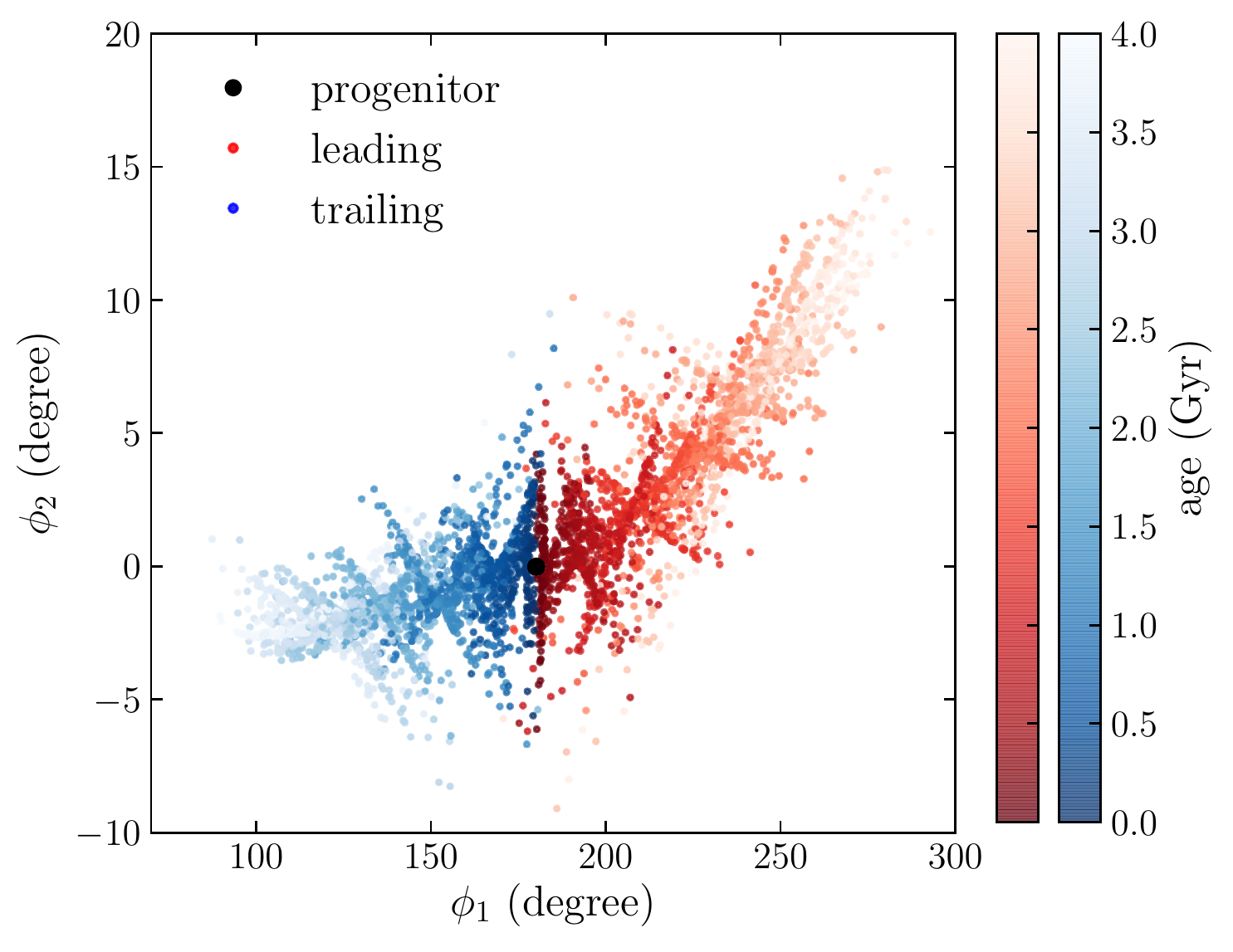}
    \caption{Simulated stream after 4 Gyr when it is still bound to its parent dwarf galaxy. The stream is shown as seen from the center of the dwarf.}
    \label{fig:stream rel sat}
\end{figure}

Figure \ref{fig:stream rel sat} displays an example stream as seen from the center of its parent dwarf galaxy $1\,\mathrm{Gyr}$ before the dwarf's merger. At this moment, the dynamical evolution of stars in the stream is still predominately determined by the potential from their parent galaxy. This simulation is based on the orbit with $t_\mathrm{merger} = 7.32\, \mathrm{Gyr}$ in Figure \ref{fig:sc orbit}, where we initiate the globular cluster $4\,\mathrm{kpc}$ from the center of the Fornax galaxy. The stream is particularly wide and long, because we are observing it at a very close distance. The colorbar indicates that the progenitor is surrounded by the stars stripped recently, while the first stars to escape are distributed at the edges. Both leading and trailing arms contain several layers of stars due to the details of the particle-spray method, but this is only noticeable because we are looking at a close-up of the stream. Other than this, the pre-merger stream does not have any significant features.

\subsection{Final stream alignment}

\begin{figure*}
    \begin{subfigure}{5cm}
    \centering
    \includegraphics[width=5cm]{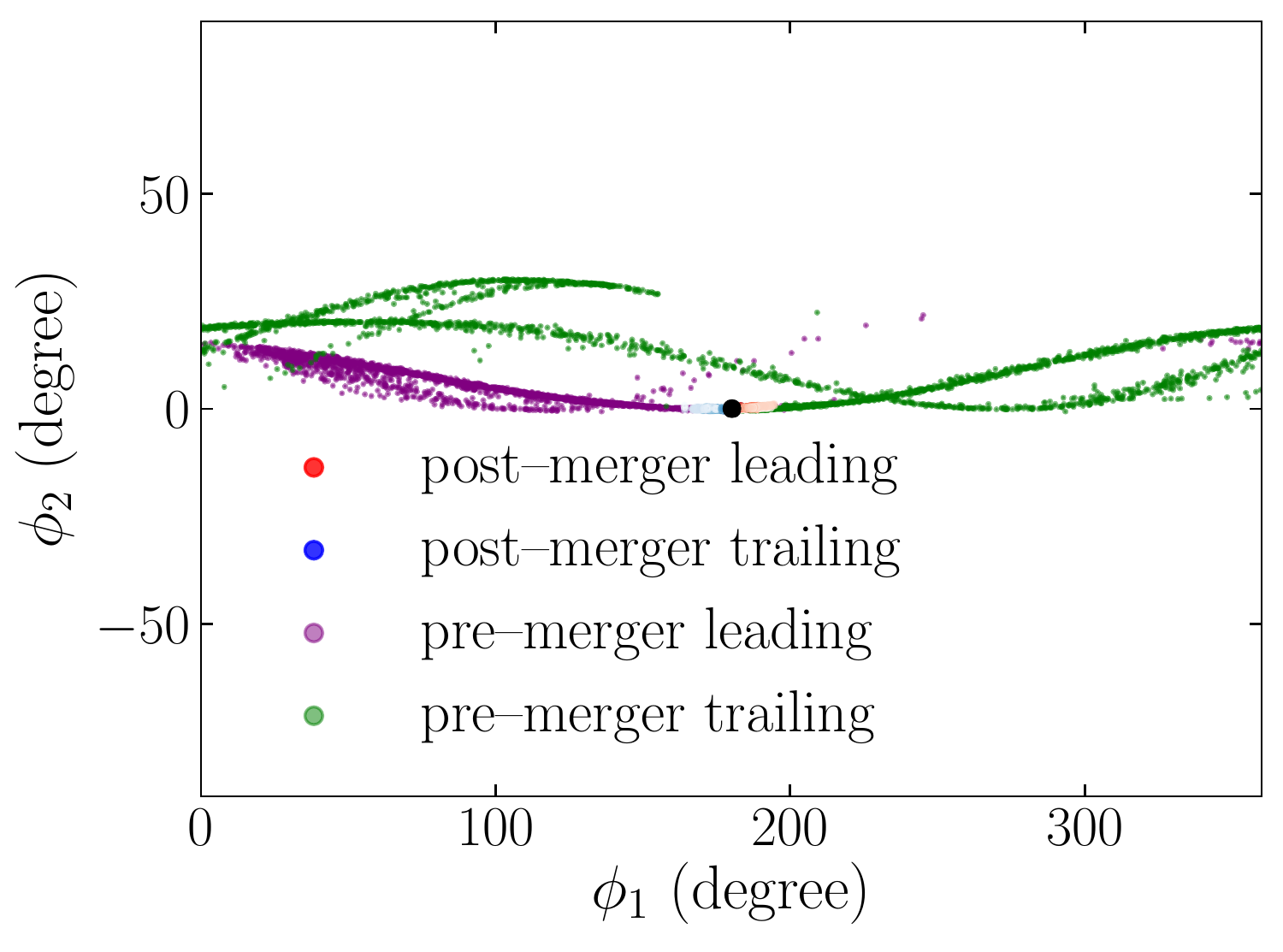}
    \caption{Fornax, approached $37\,  \mathrm{kpc}$}
     \end{subfigure}
    \begin{subfigure}{5cm}
    \centering
    \includegraphics[width=5cm]{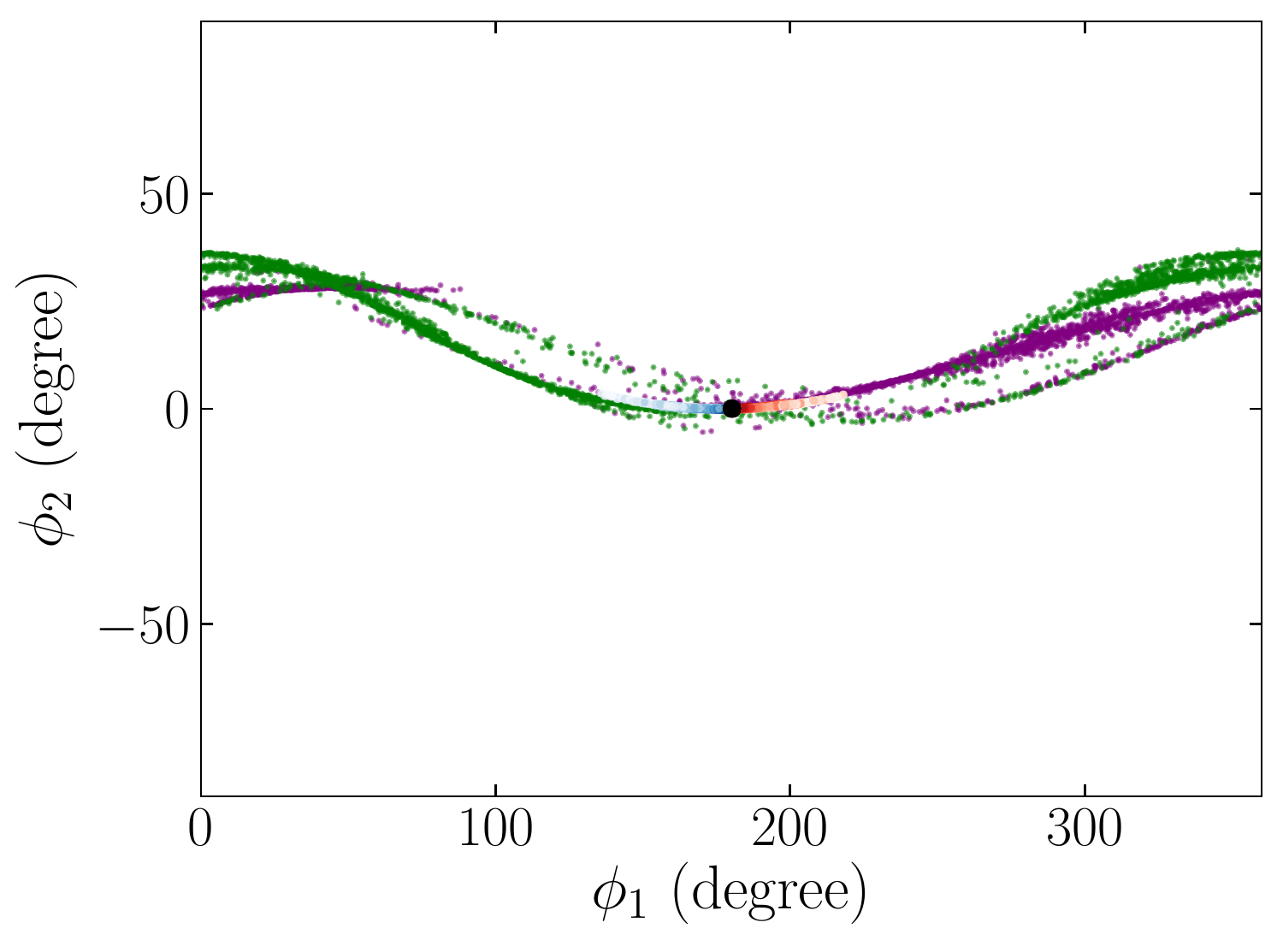}
    \caption{Fornax, approached $29\,  \mathrm{kpc}$}
     \end{subfigure}
    \begin{subfigure}{5cm}
    \centering
    \includegraphics[width=5cm]{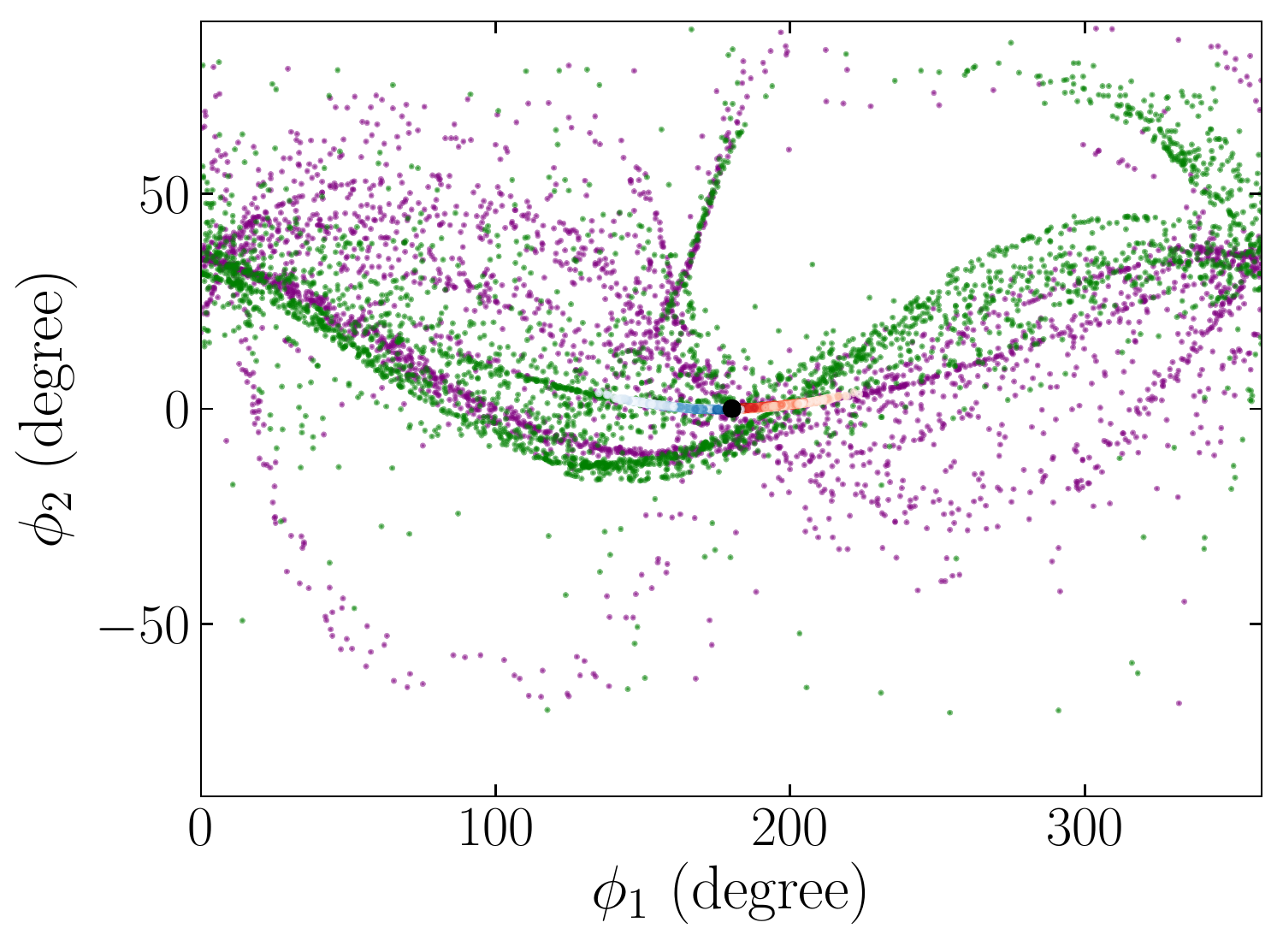}
    \caption{Fornax, approached $19\,  \mathrm{kpc}$}
     \end{subfigure}
     \\
     
     \begin{subfigure}{5cm}
    \centering
    \includegraphics[width=5cm]{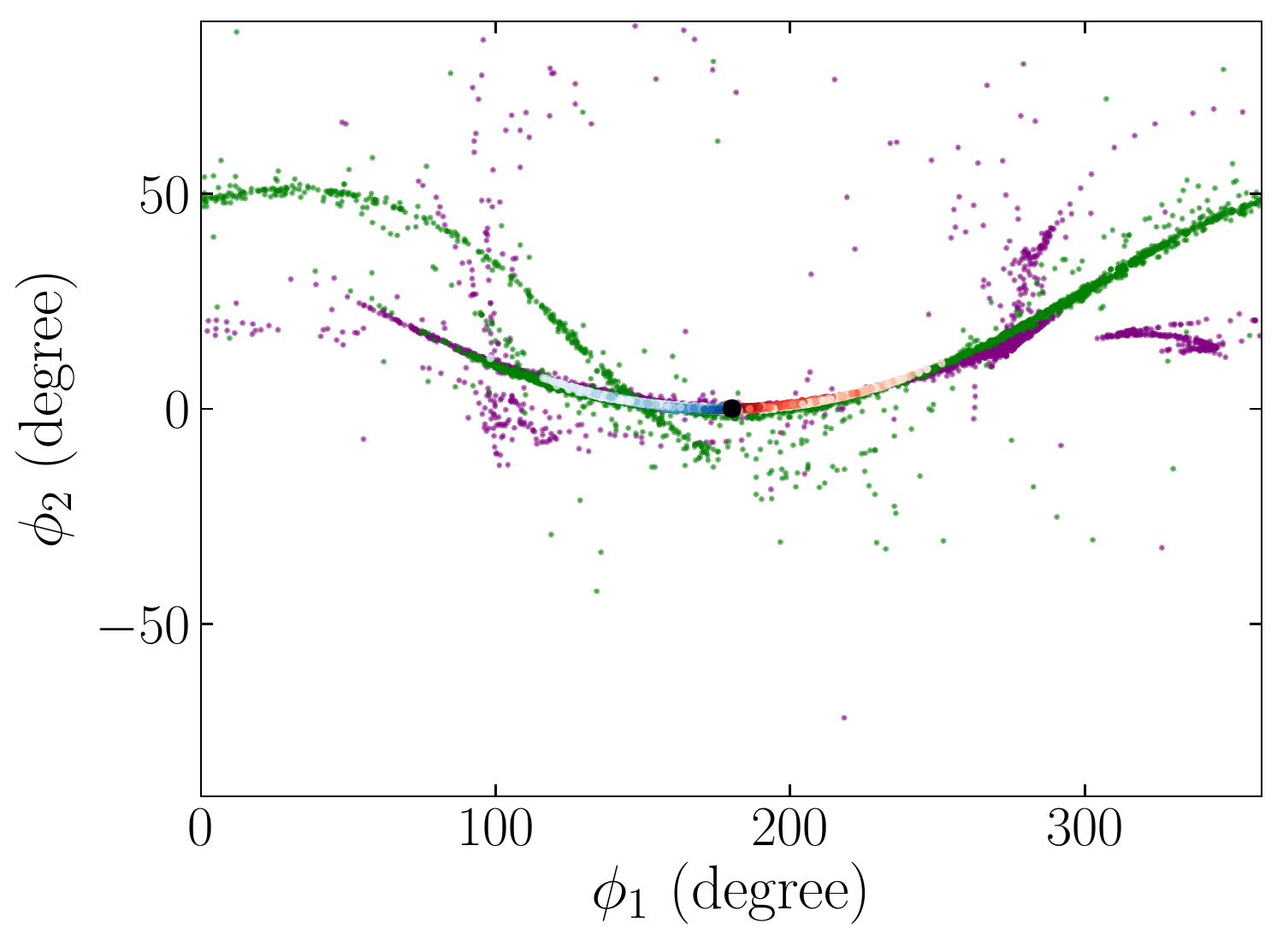}
    \caption{Draco, approached $11\, \mathrm{kpc}$}
     \end{subfigure}
    \begin{subfigure}{5cm}
    \centering
    \includegraphics[width=5cm]{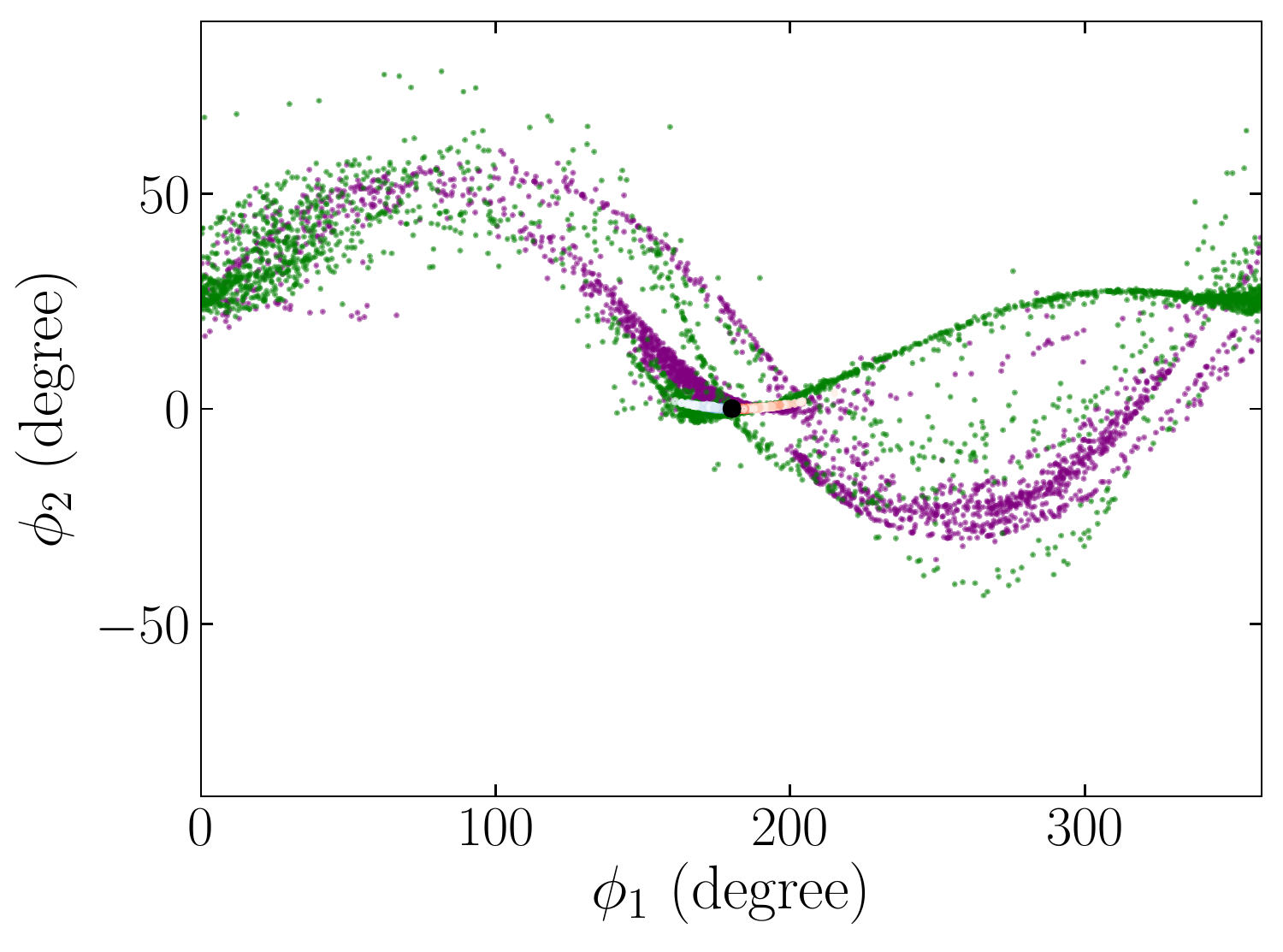}
     \caption{Draco, approached $15\, \mathrm{kpc}$}
     \end{subfigure}
    \begin{subfigure}{5cm}
    \centering
    \includegraphics[width=5cm]{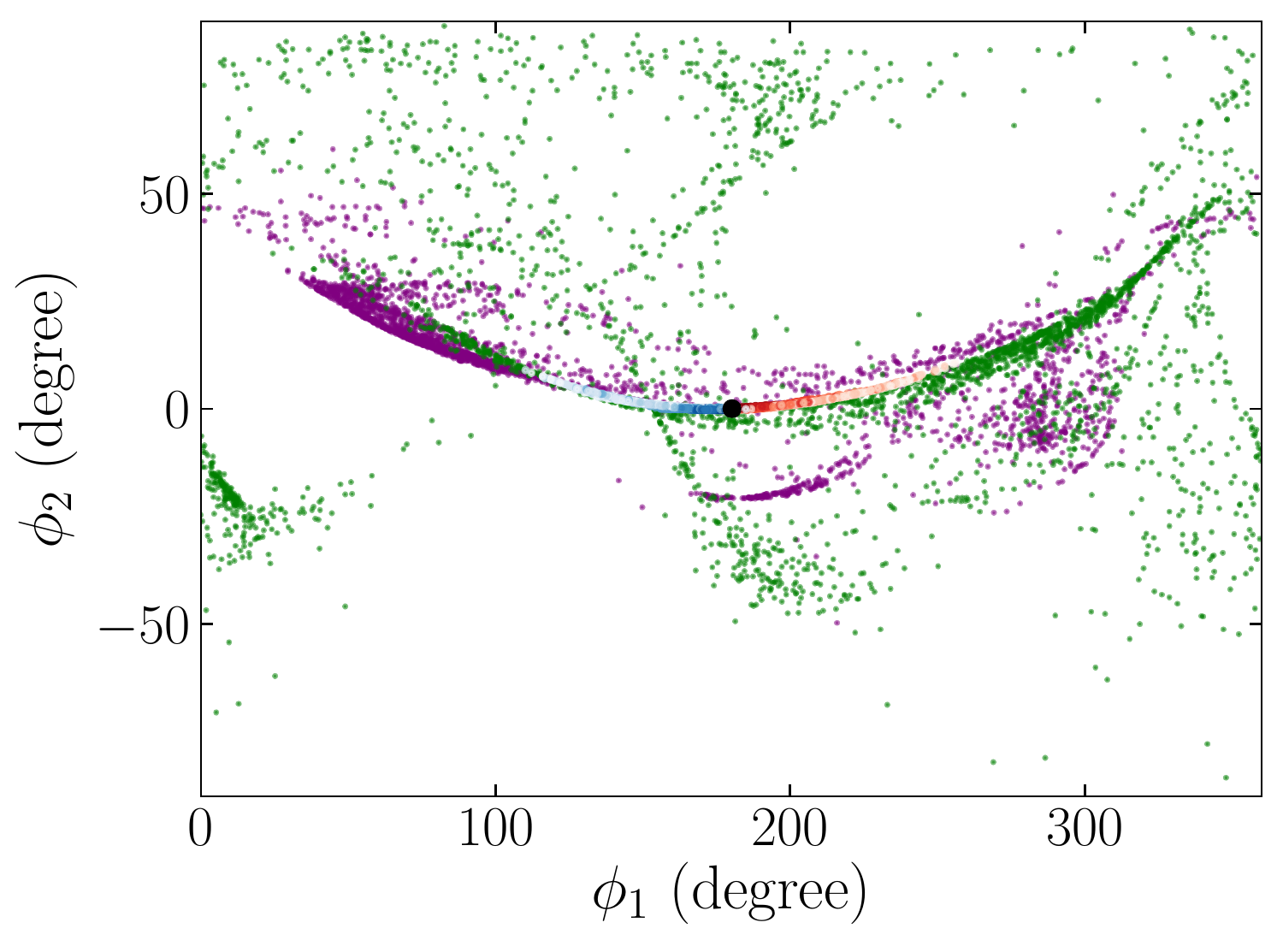}
     \caption{Draco, approached $15\, \mathrm{kpc}$}
     \end{subfigure}
     \\
     
     \begin{subfigure}{5cm}
    \centering
    \includegraphics[width=5cm]{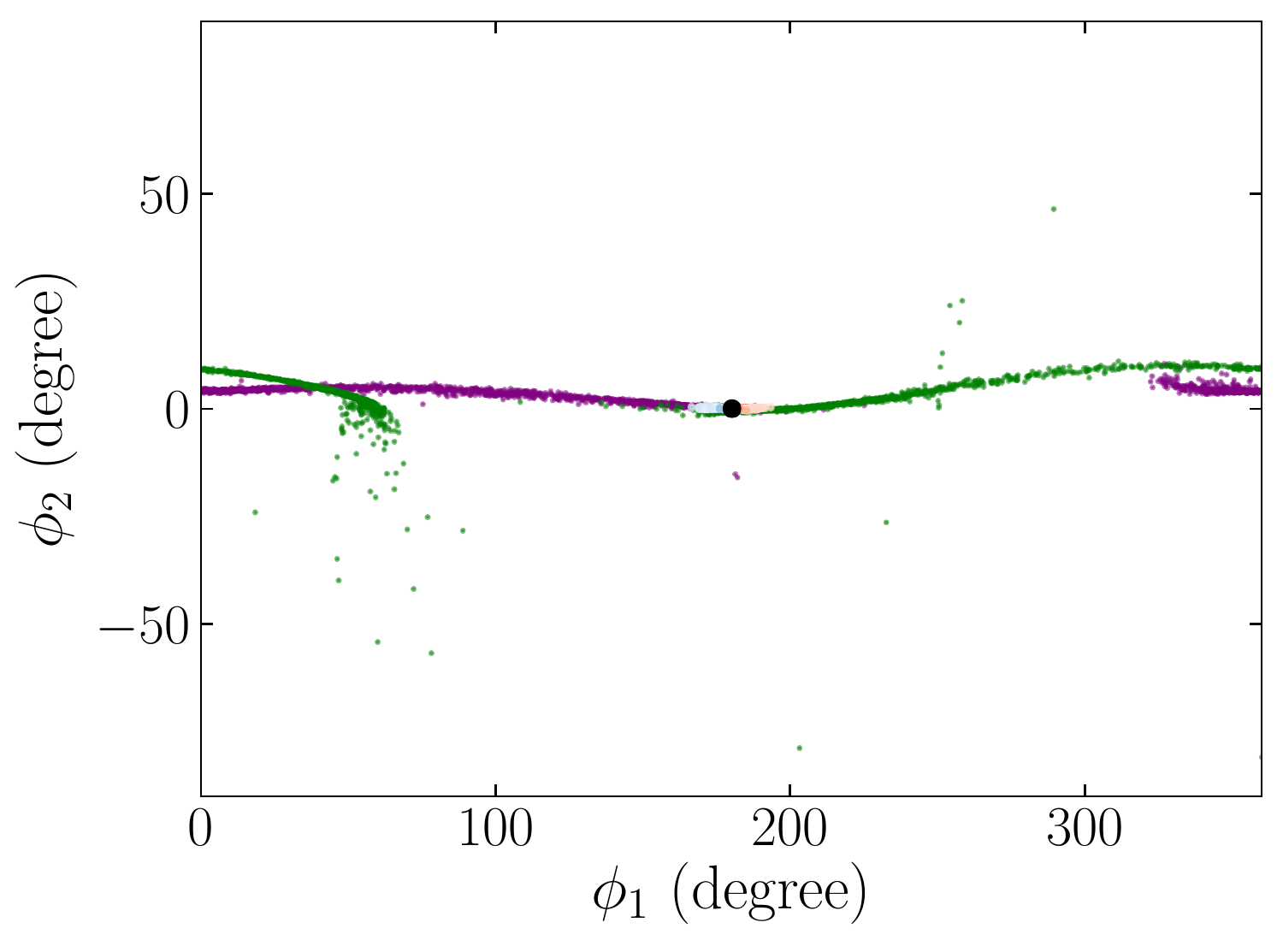}
    \caption{Sextans, approached $21\, \mathrm{kpc}$}
     \end{subfigure}
    \begin{subfigure}{5cm}
    \centering
    \includegraphics[width=5cm]{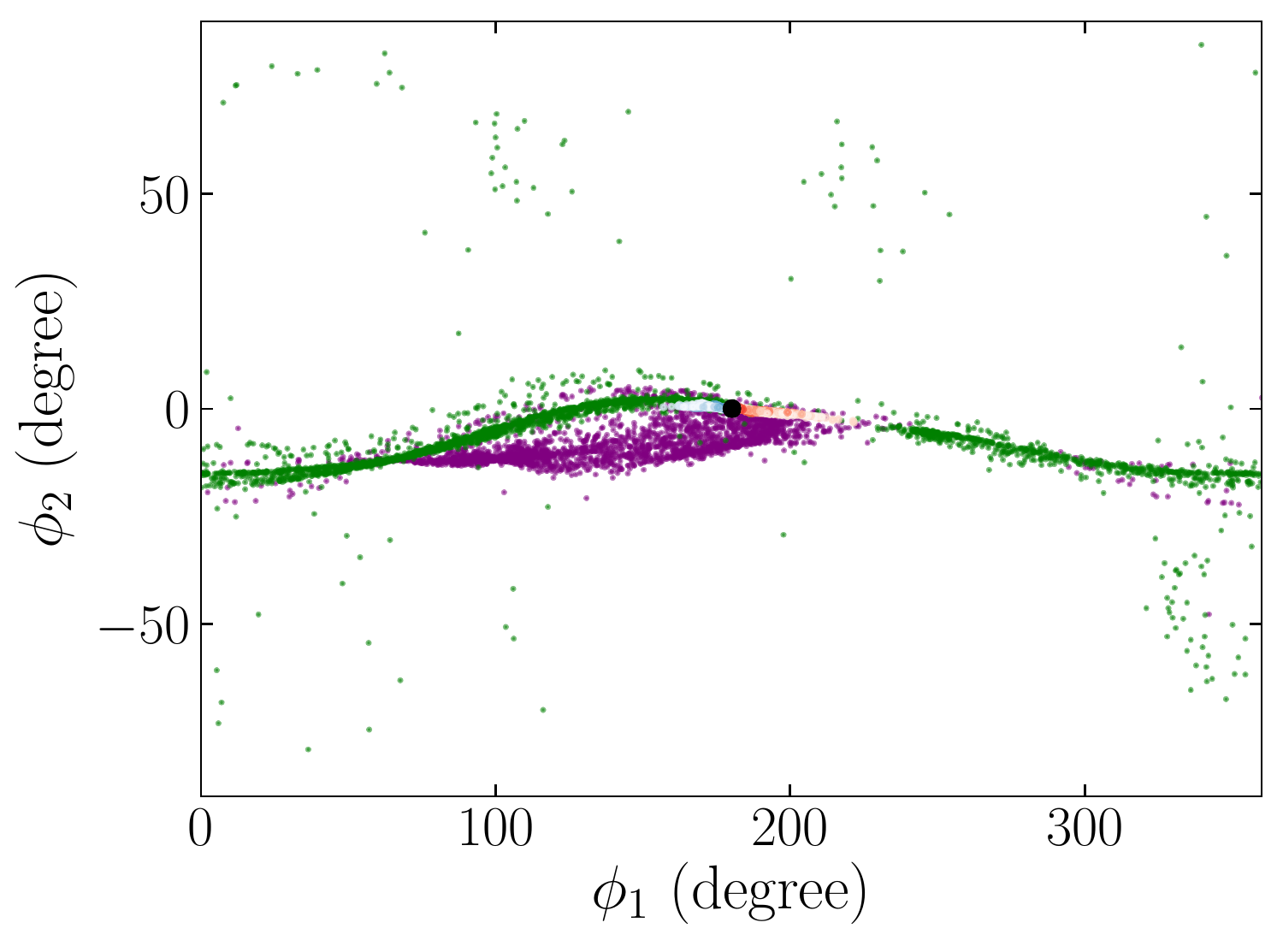}
     \caption{Sextans, approached $14\, \mathrm{kpc}$}
     \end{subfigure}
    \begin{subfigure}{5cm}
    \centering
    \includegraphics[width=5cm]{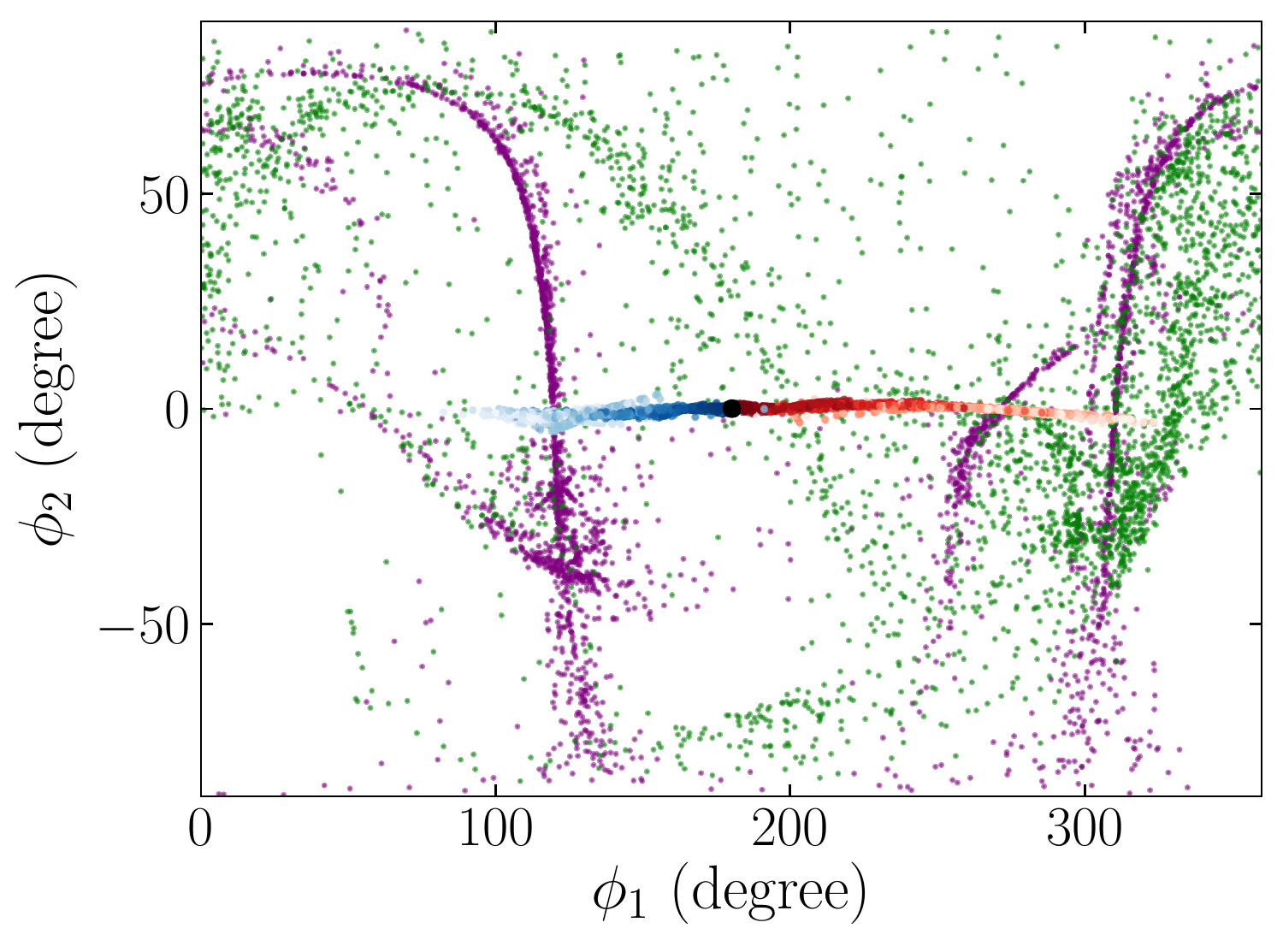}
    \caption{Sextans, approached $9\, \mathrm{kpc}$}
     \end{subfigure}
     \\
     
     \begin{subfigure}{5cm}
    \centering
    \includegraphics[width=5cm]{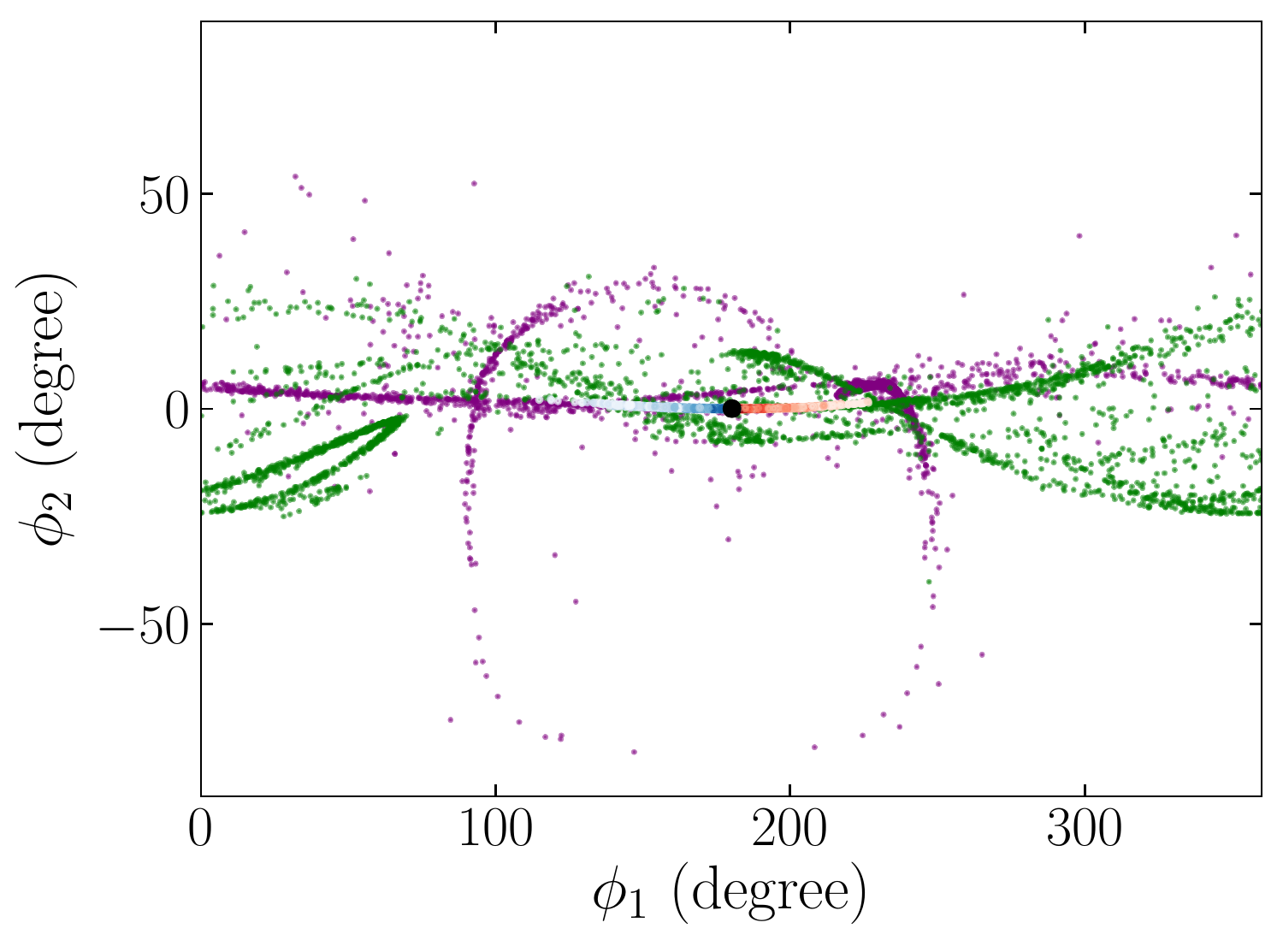}
    \caption{Willman1, approached $17\, \mathrm{kpc}$}
     \end{subfigure}
    \begin{subfigure}{5cm}
    \centering
    \includegraphics[width=5cm]{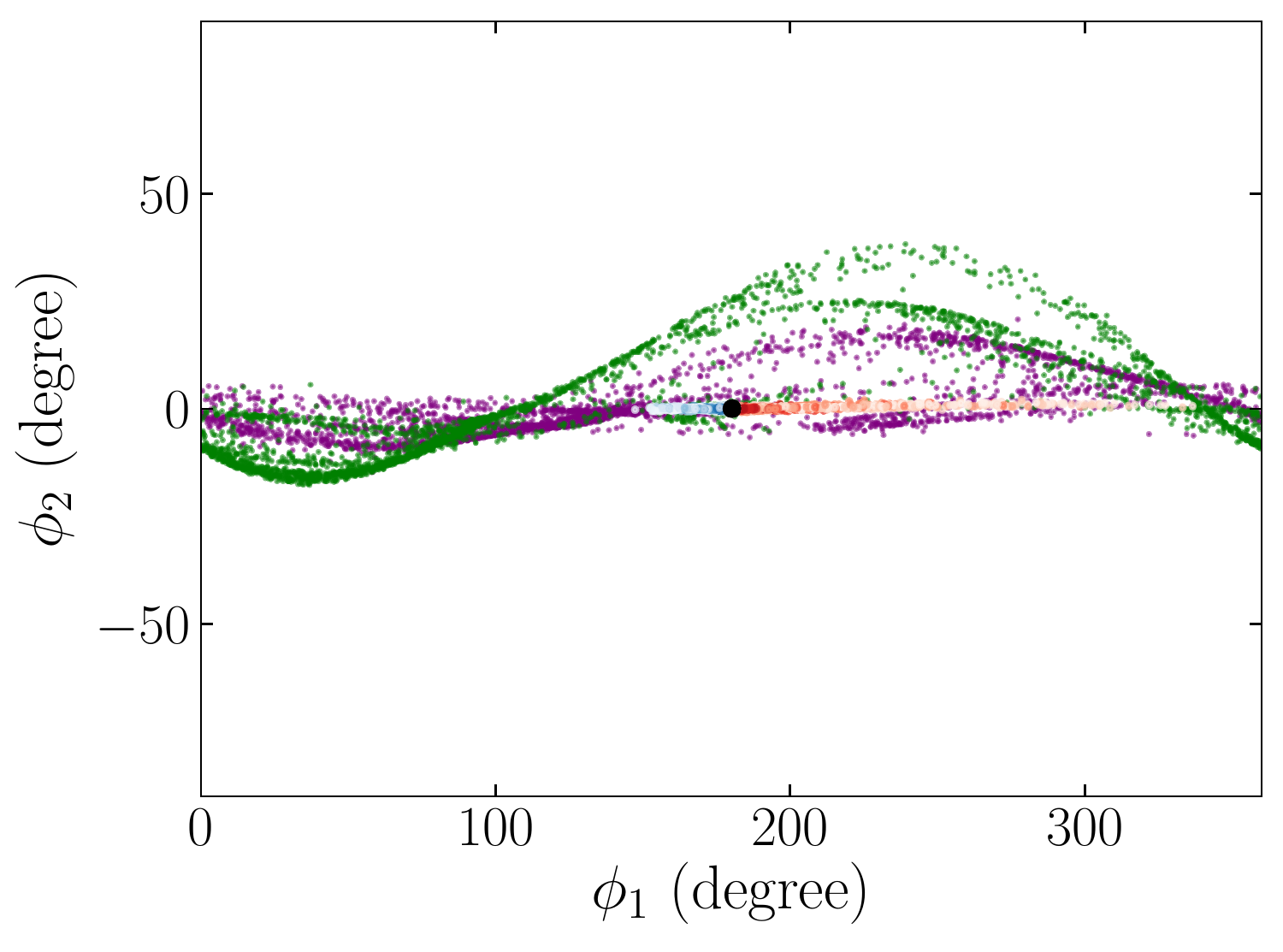}
    \caption{Willman1, approached $22\, \mathrm{kpc}$}
     \end{subfigure}
    \begin{subfigure}{5cm}
    \centering
    \includegraphics[width=5cm]{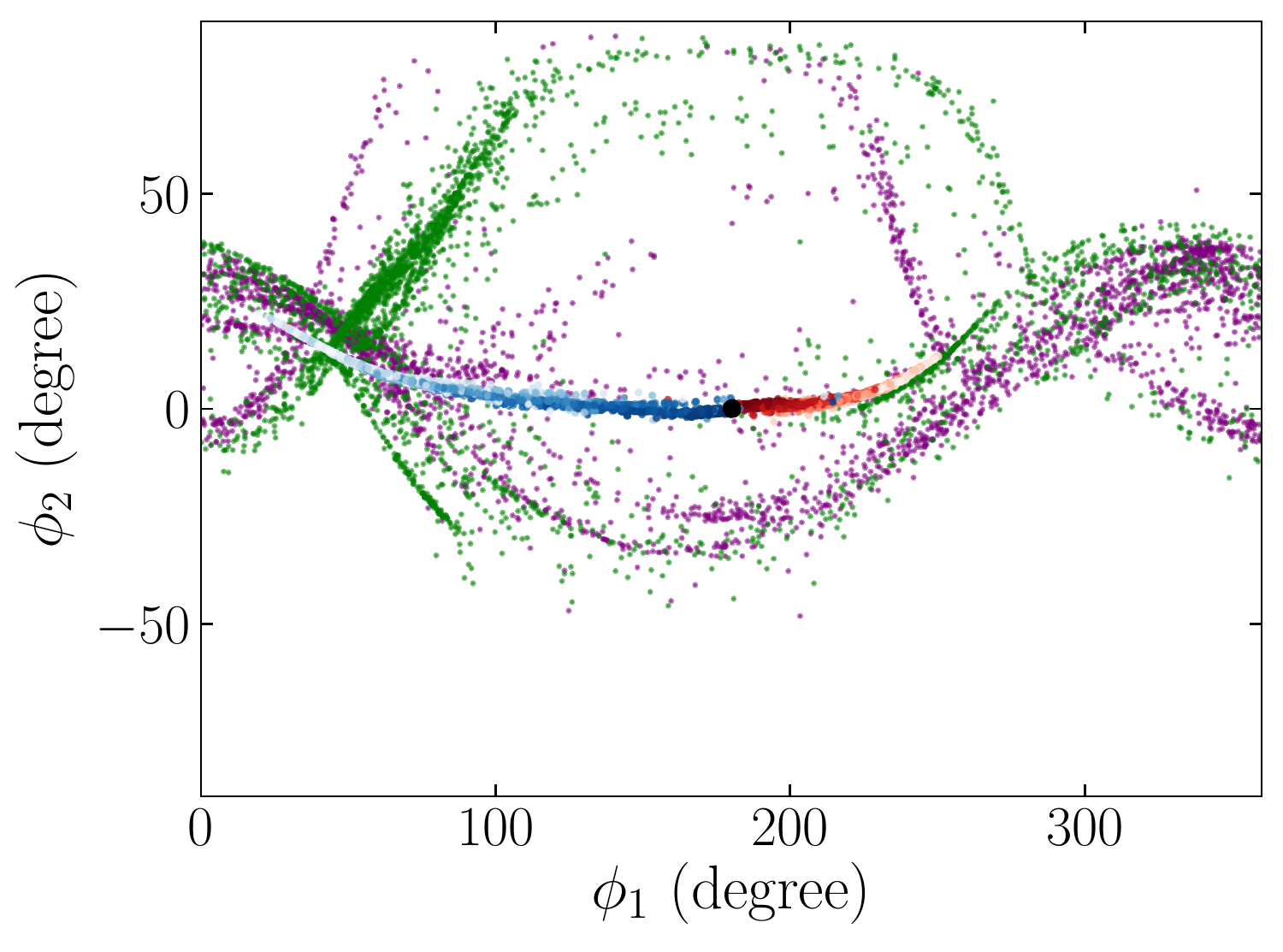}
    \caption{Willman1, approached $14\, \mathrm{kpc}$}
     \end{subfigure}
     \\
    
     \caption{Representative simulation results from different parent satellite galaxy orbits, plotted on the sky as seen from the Sun and labeled by the satellite's closest approach to the Milky Way. Purple and green stars are generated before the satellite merger event in which the globular cluster becomes bound to the Milky Way, while the graduated red and blue stars are stripped after the merger. Panels (a), (d), (g) and (j) are simulations using $4\,\mathrm{kpc}$ as the initial separation between globular cluster and satellite galaxy. The remaining panels use $2\,\mathrm{kpc}$. The accreted stream of pre-merger stars forms a network of sub-streams that can extend over a wide area of the sky depending on how close to the center the globular cluster was accreted.}
     \label{fig:res basic}
    \end{figure*}

To display and analyze the generated streams, we align them in celestial coordinate frame such that they appear approximately horizontally (similar to how this is done for observed streams). For each simulation, we calculate the relative angular momentum of the progenitor cluster to the Sun and we construct a rotational matrix that rotates this angular momentum to be in the $z$ direction. Because the stream is approximately parallel to the velocity of the progenitor cluster (as shown in figure \ref{fig:res basic}), rotating all stars by this matrix ensures that much of the stream is aligned horizontally in celestial coordinates.

\section{Results}\label{sec:results}

We simulate many different globular-cluster accretion histories using the orbits of the four different parent satellite galaxies to explore the range of behavior of the tidal stream formed through the pre- and post-merger tidal stripping of the globular-cluster progenitor. To vary the history for each parent satellite orbit, we start globular clusters at either 2 kpc or 4 kpc from the center of the satellite and then disrupt/merge the satellite at different galactocentric distances to vary how close to the center the globular cluster is accreted. At the final time, the resulting tidal stream is typically within 20 kpc from the center of the Milky Way, similar to observed tidal streams. In this section, we present the results from this analysis.

\subsection{A wide range of accreted-stream behavior}

\begin{figure*}
    \centering
    \includegraphics[width=0.85\linewidth]{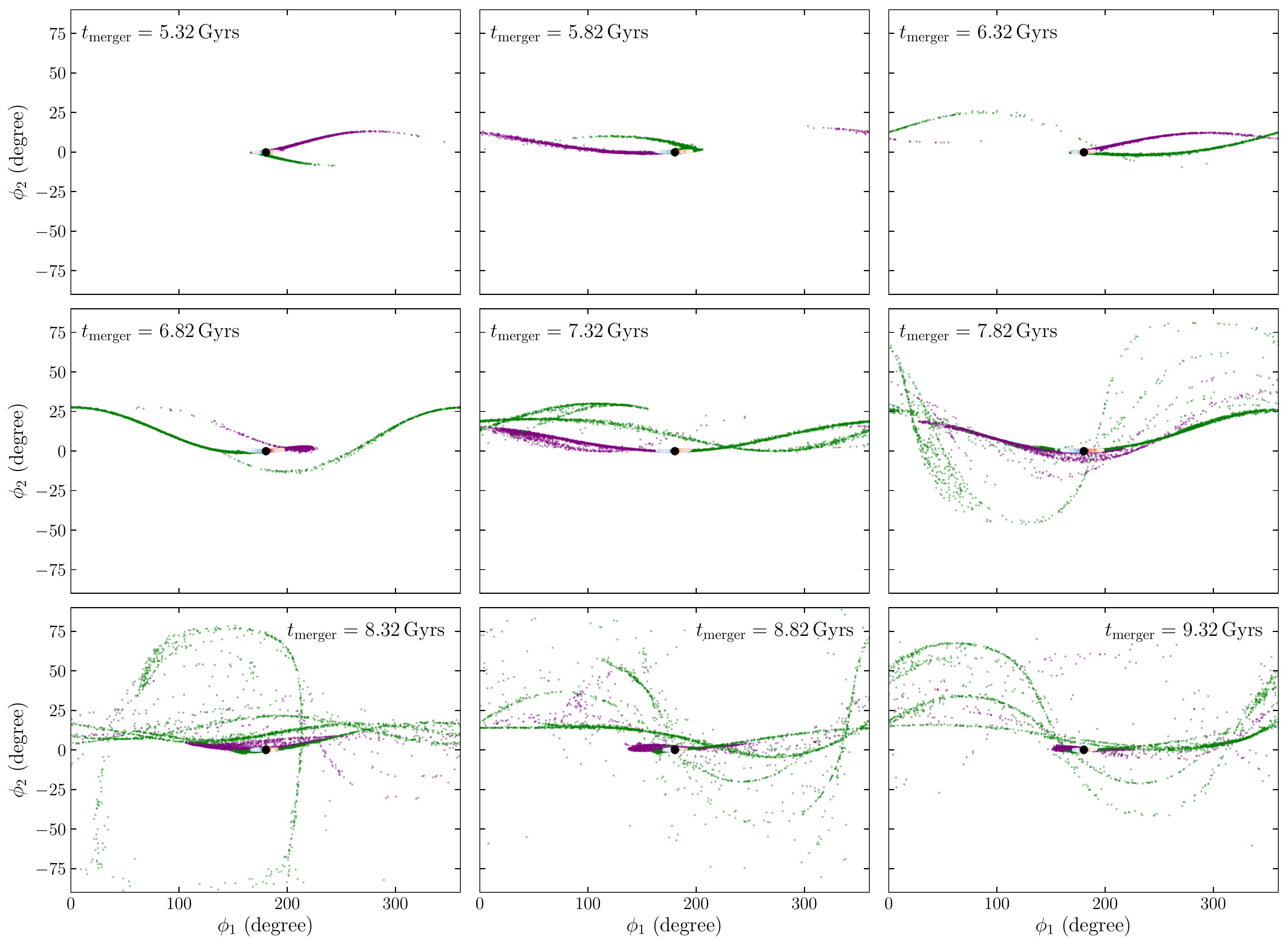}
    \caption{Effect of the time and location of the merger that unbinds the stream progenitor from its parent satellite galaxy. The different panels show streams generated along the orbit of Fornax with different merger times that correspond to different closest approaches of the satellite. Earlier mergers happen at larger distances and lead to relatively simple present-day structures of the accreted streams. As the merger happens later closert to the galactic center, the accreted stream consists of more sub-streams and fills a wider area on the sky.}
    \label{fig:tshift}
\end{figure*}

Figure \ref{fig:res basic} gives twelve example accreted streams, three for each parent-satellite orbit. We have chosen three simulations per parent satellite from among a dozen simulations that represent the range of behavior seen in the larger set of simulations. By combining the globular-cluster radius (2 or 4 kpc) and time of the satellite merger, we can choose how close the parent satellite gets to the center of the Milky Way before it merges with the Milky Way and the globular cluster becomes a satellite of the Milky Way itself; this closest approach is indicated in the figure. 

As expected from the theory of tidal-stream formation in a smooth, static potential along an orbit in that potential, the tidal debris from the post-merger stars---stars tidally stripped after the globular cluster is unbound from its parent satellite galaxy---forms thin leading and trailing streams emerging from the progenitor \citep[e.g.,][]{Bovy2014}. Each of the leading or trailing arms extends over 10 to 30 degree from the progenitor depending on the stream's distance from the Sun and its orbital phase. The pre-merger stars---stars tidally stripped before the globular cluster becomes unbound from its parent satellite---are, however, distributed over a much larger area of the sky. The pre-merger stars usually form multiple long streams that are less dense than the post-merger stream around the progenitor. We refer to these structures as ``sub-streams'' to distinguish them from the main post-merger stream. They are usually parallel to the main stream, with offsets ranging from a few degree to more than $20$ degree above or below. 

Figure \ref{fig:res basic} demonstrates that depending on how close to the center of the Milky Way the globular cluster was stripped from its parent satellite, the accreted stream of pre-merger stars has a range of morphologies. If the accretion happens at large distances $\gtrsim 20\,\mathrm{kpc}$. the structure of the sub-streams is typically relatively clean, with a few dense sub-streams dominating the accreted stellar stream. Thus, in this case each of the sub-streams could show up as its own stellar stream. In some cases, the sub-streams are so close to the post-merger stream that they almost merge with it and, thus, show up as a widening of the post-merger stream. Globular clusters that are accreted at smaller distances from the center, however, display a much more erratic accreted stellar stream. While we can still distinguish sub-streams, they are wider and a significant number of stars are distributed between the sub-streams, forming a two-dimensional locus on the sky. Part of this is a projection effect, because these streams are typically observed somewhat closer by, but much of it is a true physical wider debris distribution. Because of its low density, this structure would be very difficult to observe.

\subsection{Effect of the satellite's closest approach}

To further investigate the importance of how close to the Milky Way's center that a globular cluster separated from its parent satellite (what we refer to as ``the merger''), we take one of our simulations using the orbit of Fornax and unbind the globular cluster from its parent satellite at different times (these are different times at which we let the mass of the satellite decay to zero). The fiducial simulation in the set lets the satellite's mass decay as soon as it enters within 40 kpc from the center and the other simulations move this merger time forward or backward by 2 Gyr. The effect of this change on the orbit of the accreted globular cluster in the Milky Way is shown in Figure \ref{fig:sc orbit}. Because the closest approach of the parent satellite to the Milky Way is at a smaller distance for a later accretion time due to the effect of dynamical friction, the galactocentric distance of the globular cluster and its stellar stream today  is roughly inversely proportional to the merger time. 

The effect of changing the merger time on the resulting tidal streams today is shown in Figure \ref{fig:tshift}. In this figure, the stream that becomes unbound from its parent dwarf galaxy early at $5.32\, \mathrm{Gyr}$ only consists of a main stream, and the pre-merger stars (purple and green dots) still tend to locate around the globular cluster. As the merger time increases, the pre-merger stars start to spread out. At some point, some stars are one cycle ahead of the main stream and start forming sub-streams. More sub-streams are formed, and they start to occupy the entire sky if we delay the merger further. At the same time, the angle between the sub-stream and main stream becomes more steep, reaching nearly $90$ degrees for the stream that merged at $8.32\, \mathrm{Gyr}$. These results confirm those from the previous section: accreted streams from globular clusters accreted in the outskirts of the Milky Way are relatively simple, consisting of a few sub-streams at most, while those of globular clusters accreted closer to the center can display complex behavior.

\subsection{Action-angle coordinates} \label{sec action}

\begin{figure*}
    \begin{subfigure}{6cm}
    \centering
    \includegraphics[width=6cm]{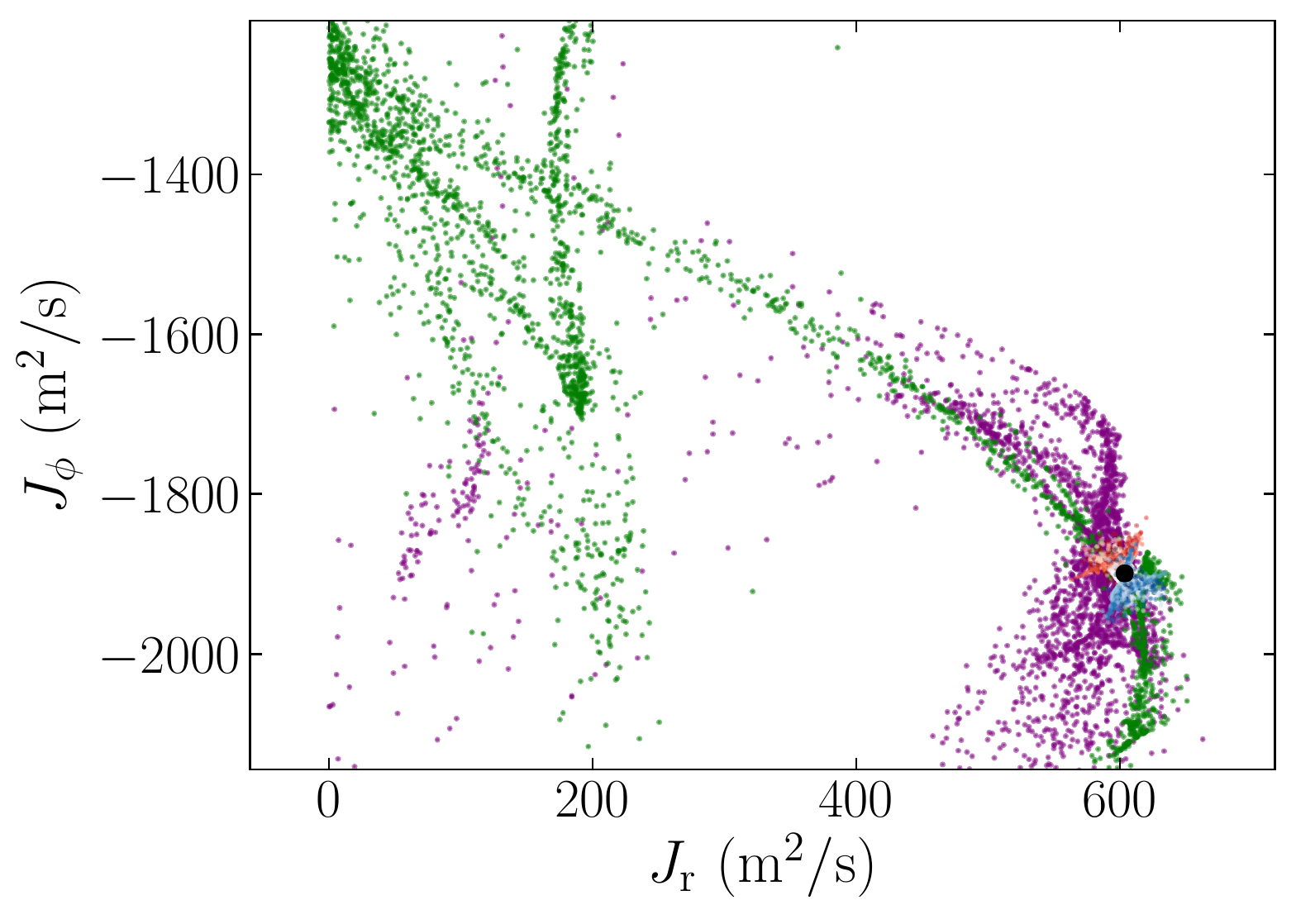}
     \end{subfigure}
     \hspace{1cm}
    \begin{subfigure}{6.3cm}
    \centering
    \includegraphics[width=6.3cm]{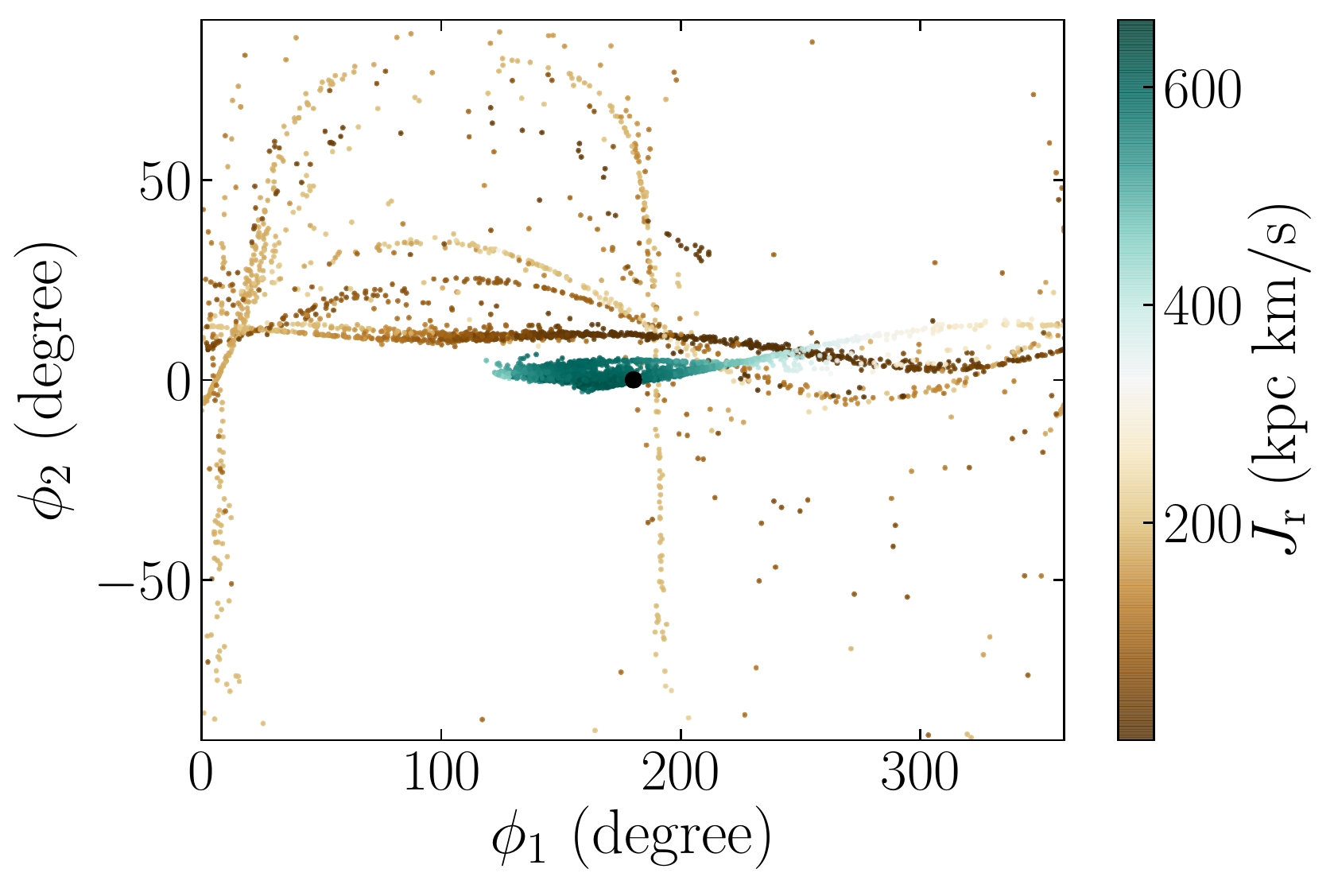}
     \end{subfigure}
     \caption{Action distribution (left panel) of a simulated stream that displays multiple sub-streams on the sky (right panel). The sky distribution is color-corded by the radial action, which has very different values for the different sub-streams.}
     \label{fig:largejr}
\end{figure*}

\begin{figure*}
    \begin{subfigure}{6cm}
    \centering
    \includegraphics[width=6cm]{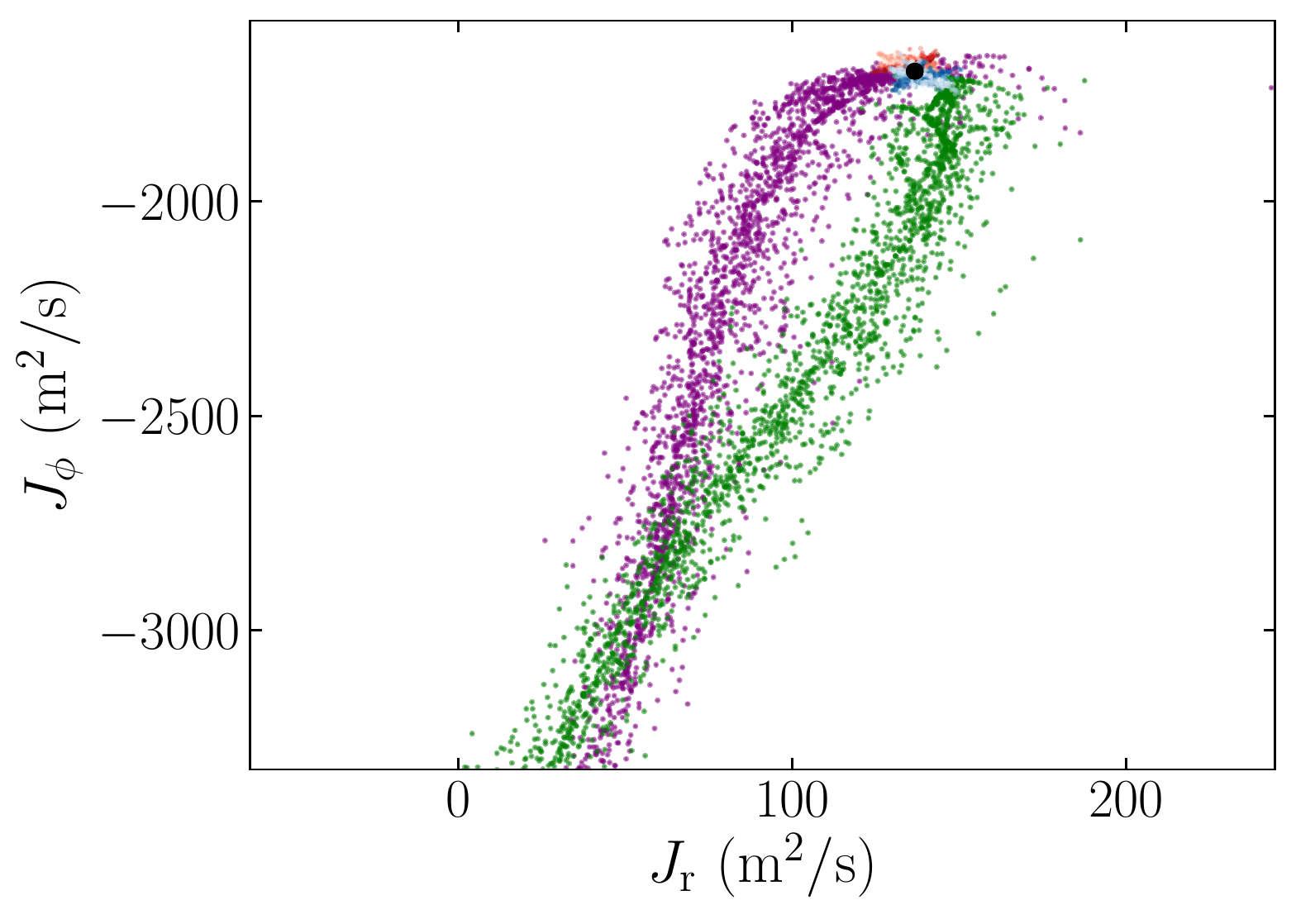}
     \end{subfigure}
     \hspace{1cm}
    \begin{subfigure}{6.3cm}
    \centering
    \includegraphics[width=6.3cm]{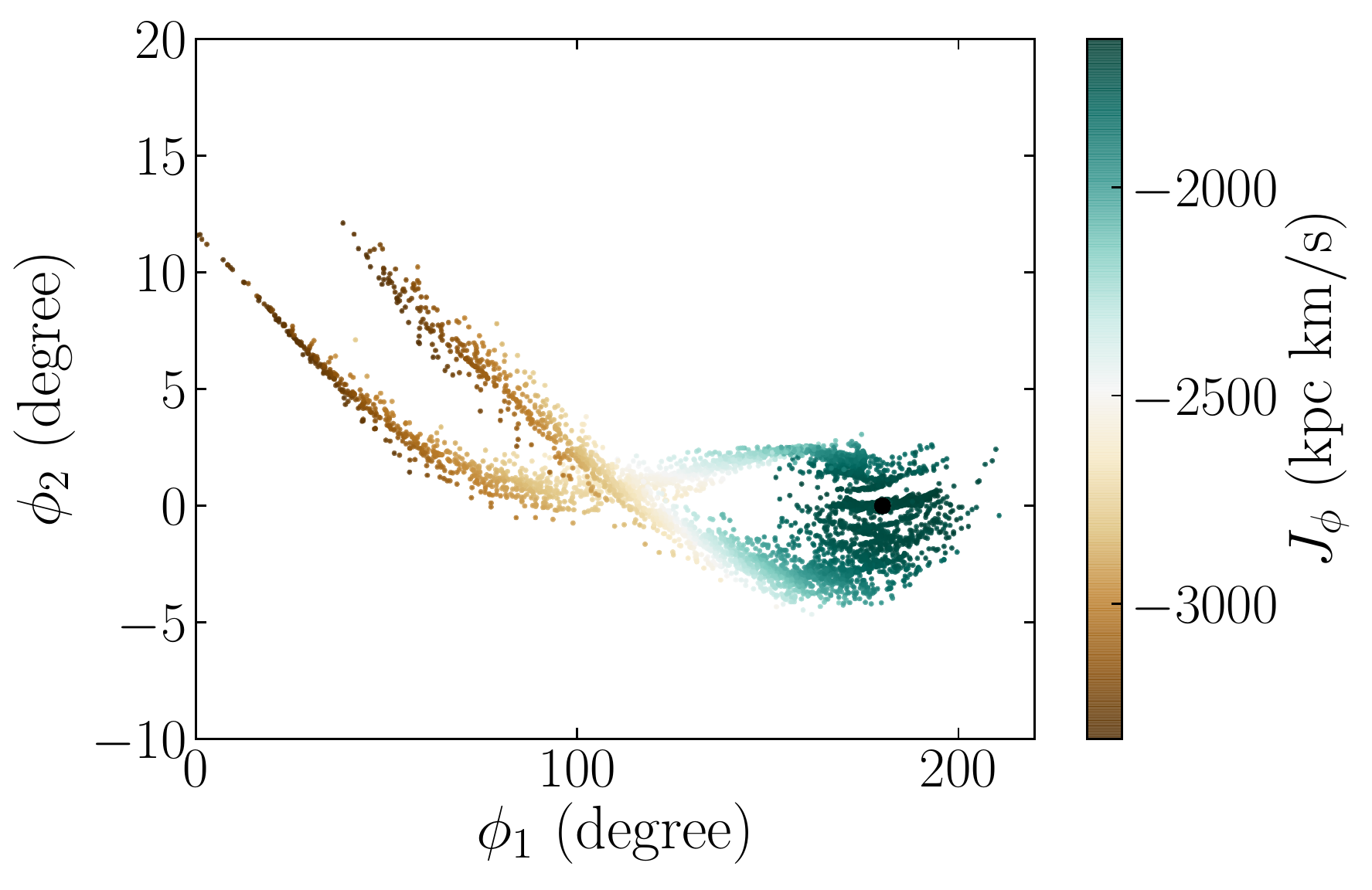}
     \end{subfigure}\\
      \begin{subfigure}{6cm}
    \centering
    \includegraphics[width=6cm]{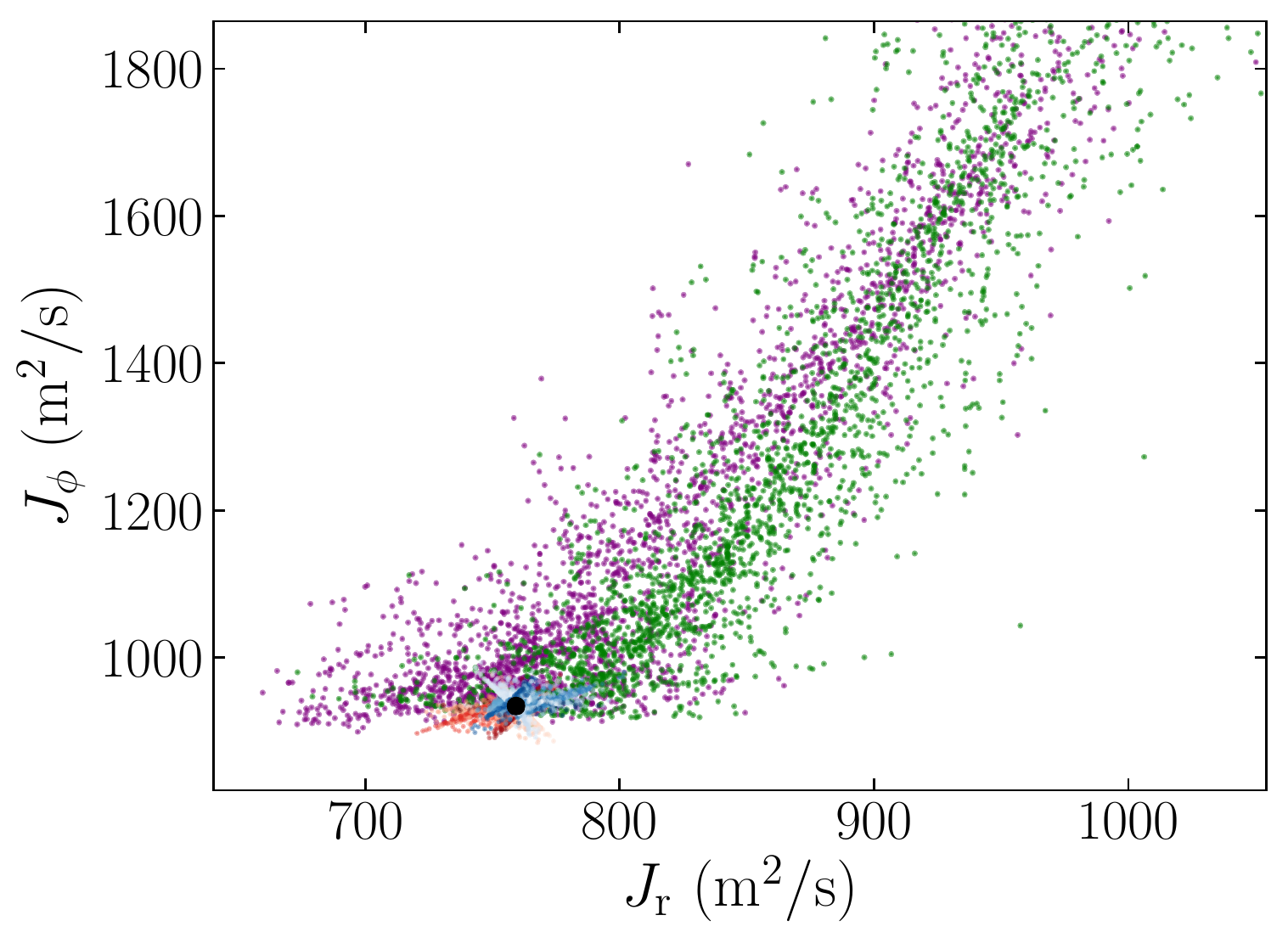}
     \end{subfigure}
     \hspace{1cm}
    \begin{subfigure}{6.3cm}
    \centering
    \includegraphics[width=6.3cm]{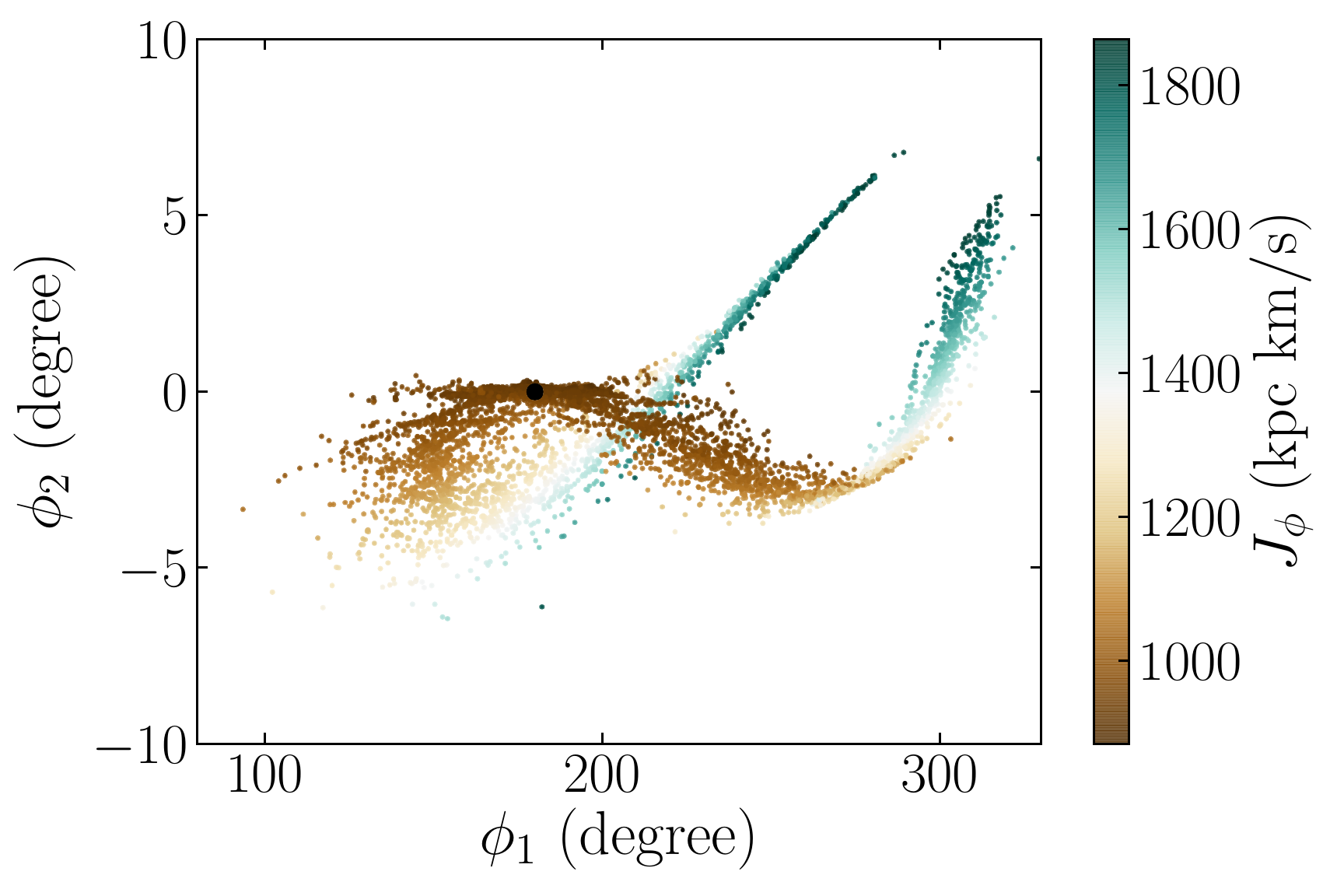}
     \end{subfigure}
     \caption{Like Figure \ref{fig:largejr}, but for simulated streams that have leading and trailing arms with a large gradient in the azimuthal action $J_\phi$.}
     \label{fig:largejp}
\end{figure*}

To investigate the origin of the structure seen in the accreted stream, we convert the stream phase-space distribution to action-angle coordinates. We do this in the static \texttt{MWPotential2014} potential using the convergent method from \citet{Bovy2014}.  Specifcally, we use the radial action $J_r$ and the azimuthal action (the angular momentum) $J_\phi$; because the stream stars orbit in regions where the potential is approximately spherical, there is little additional information in the vertical action $J_z$. Actions computed in this way are conserved for orbits in the \texttt{MWPotential2014} potential, but before the merger the stars orbit in the combined, time-dependent \texttt{MWPotential2014}+parent-satellite potential and during this stage the actions of stream stars change in time. The structure of the final stream is determined by the structure in action space that is frozen once the parent satellite disappears. Of course, the angle distribution of the stream is not uniform and correlations between actions and angles play a large role in determining the final structure.

Example action distributions are displayed in Figures \ref{fig:largejr} and \ref{fig:largejp} together with the stream's present position on the sky. As expected from tidal-stream formation theory \citep[e.g.,][]{Bovy2014}, the post-merger stream occupies a small region of action space around the location of the progenitor, with this region consisting of two clouds representing the leading and trailing arm of the stream. However, the accreted stream spans a much wider range of actions than the post-merger stream. Surprisingly, in most cases the action distribution has significant substructure and looks quite streamy itself. Figure \ref{fig:largejr} gives an example where the spread in the radial action of the accreted stream is large and the $J_r$-color-coded sky distribution demonstrates that the sub-streams are separated from each other and from the post-merger stream by large $J_r$ differences, indicating that the orbits of the sub-streams are quite different from that of the post-merger stream. Figure \ref{fig:largejp} gives a few examples where the accreted stream has a large spread in angular momentum. In this case, the leading and trailing arms tend to curve up, and the width of arms drops dramatically from the progenitor to the tail. This type of accreted stream covers a larger two-dimensional area on the sky and is as such distinguishable from the one-dimensional sub-stream case.

\begin{figure}
    \centering
    \includegraphics[width=\linewidth]{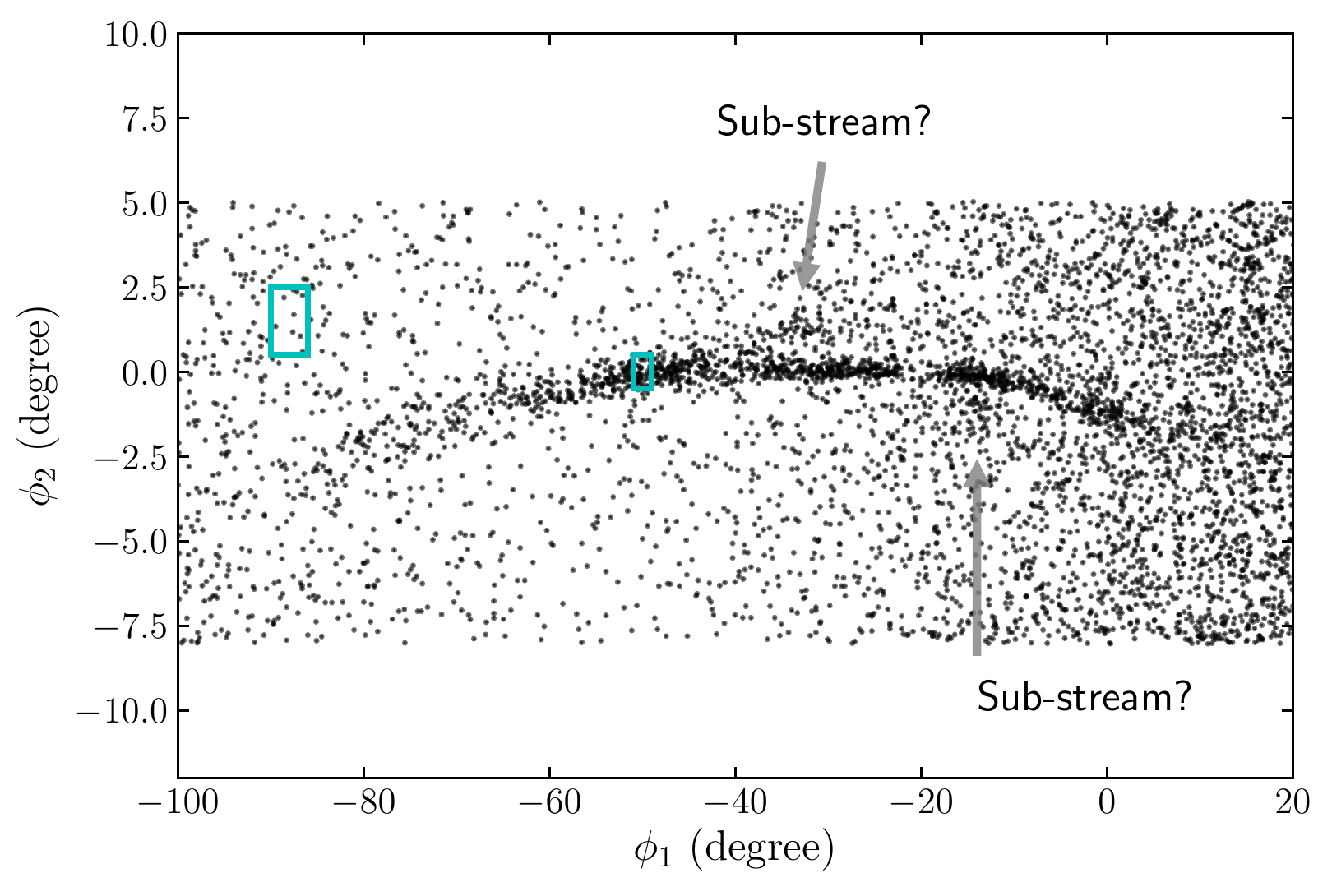}
    \caption{GD-1 stars from Gaia filtered by proper motion and photometry. The labeled off-track features may be examples of sub-streams in the accreted-stream component of GD-1. The indicated boxes are the areas used to determine the stream-to-background signal-to-noise ratio used to compare simulated streams to this observation.}
    \label{fig:gd1}
\end{figure}

\begin{figure}
    \centering
    \includegraphics[width=\linewidth]{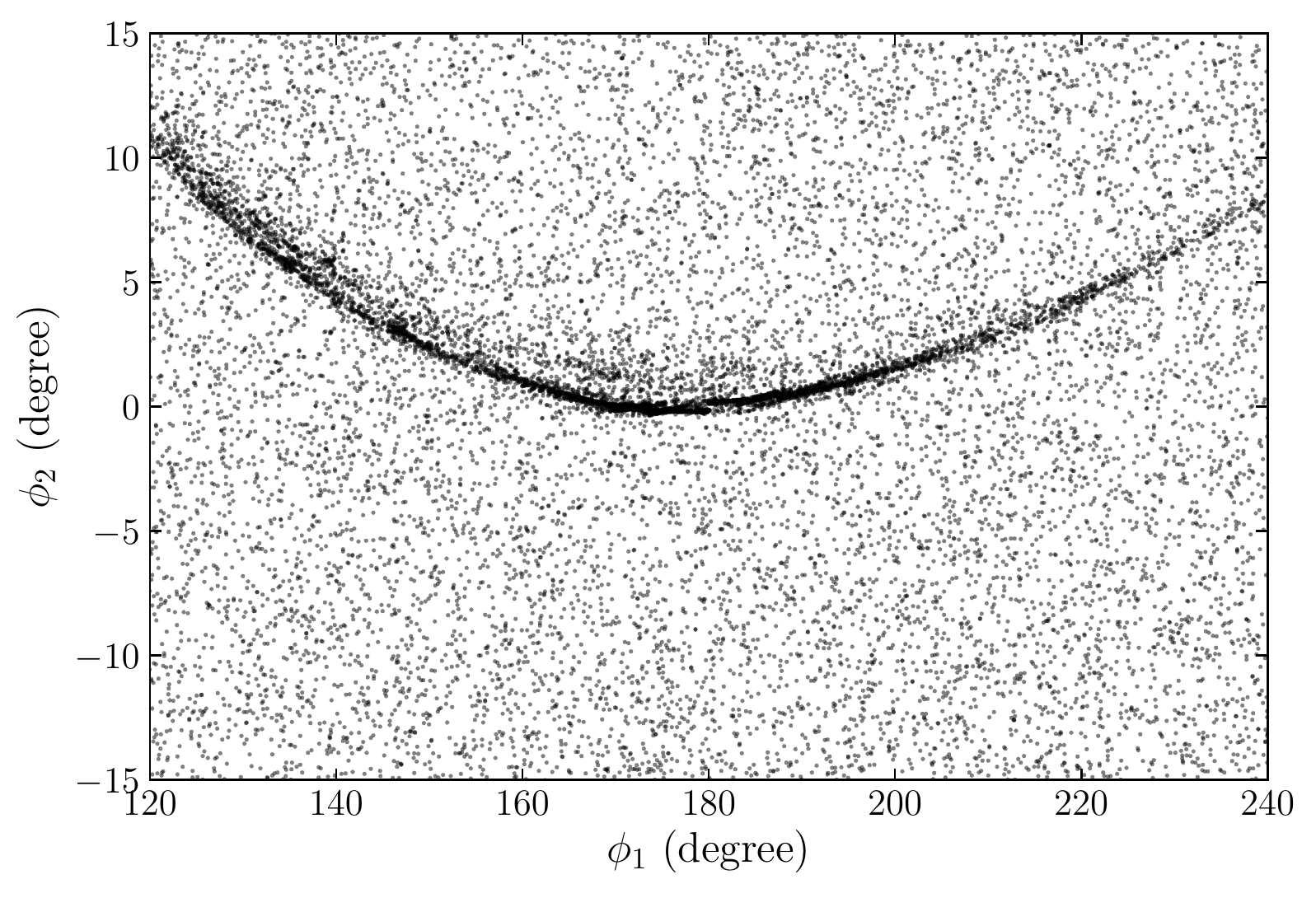}
    \caption{Example of a simulated stream that displays similar features as observed around the GD-1 stream in Figure \ref{fig:gd1}. This stream is generated from a globular cluster accreted from Draco at $2.23\,\mathrm{Gyr}$. To properly compare the simulated stream to the observations, we add a random background of stars with a similar ratio as that for the observed GD-1 stream. A sub-stream is observable $\sim 1$ degree above the main stream.}
    \label{fig:gd1comp1}
\end{figure}

Overall, the action distribution of the accreted stream is complex and displays a wide range of behavior. This is the case even for a globular cluster stripped from the same parent satellite along different phases of the latter's orbit, such as that shown in Figure \ref{fig:tshift}. The large range spanned in both actions results both from the mass of the parent satellite, which causes the action distribution to be wide already before the merger, and the randomness of the process that strips the globular cluster and its accreted stream from the parent satellite. The closer the parent satellite gets to the center of the host galaxy before merging, the wider the range of actions in the accreted streams becomes, and the more sub-streams and other structure occur, as clearly demonstrated in Figure \ref{fig:tshift}.

\section{Comparison to observations}\label{sec:observations}

The Gaia mission \citep{GaiaMission} with its exquisite five-dimensional kinematic data and photometric data, has made the discovery of many new streams and of many new features of known streams possible \citep[e.g.,][]{streamfinder,price2018,Palau2019,Starkman2020,ibata2020}. Aside from finding the archetypical cold streams of which many examples were already known before Gaia, the Gaia data in combination with auxiliary data have revealed additional structure associated with known stellar streams, in particular around the GD-1 stream \citep{Grillmair2006}. These additional structures occur near the main track of the stream, but are clearly separated from it as they extend out to a degree or more, while the main stream is well within a degree. These structures are sometimes referred to as a ``cocoon'' \citep{Malhan2019} or ``spur'' or ``blob'' \citep{price2018}, although these do not refer to the exact same structure (``cocoon'' means diffuse structure while ``spur'' and ``blob'' demonstrate the off track nature). As suggested by \citet{Malhan2019}, the cocoon structure could be the effect of an accreted stream and in this section we employ the simulations that we presented in the previous section to investigate whether accreted streams can be more generally responsible for cocoons, spurs, or blobs. Because their data is readily available, we focus on the spur and the blob as defined by \citet{price2018} here.

In Figure \ref{fig:gd1}, we use the data from \citet{price2018} to plot the GD-1 stream as observed by Gaia. These authors filter the GD-1 stars by their proper motion obtained using Gaia and their photometry from the Pan-STARRS survey \citep{gaia, chambers2016}. They assume all GD-1 member stars belong to a single population and use an isochrone corresponding to an old, metal-poor stellar population to select main-sequence stars of a given age and metallicity. Two substructures can be identified in Figure \ref{fig:gd1} separated more than 1 degree from the main stream: the spur, which is above the main stream at $\phi_1 \approx -40^\circ$ and the blob, which is below the stream at $\phi_1 \approx -20^\circ$. The question is then whether these can be reproduced by the sub-stream features found in our simulations of accreted streams.

To compare our simulations to the observations, we add a background of stars by approximately matching the signal to noise ratio (SNR) in Figure \ref{fig:gd1} (note that many of the non-stream stars in Figure \ref{fig:gd1} are in fact foreground stars, but the distinction does not matter here). The SNR is taken by the ratio of the density of stars in the GD-1 stream, estimated using the boxes along the main stream (like the one in Figure \ref{fig:gd1} at $\phi_1 \approx -50^\circ$), to those in the background, estimated to the off-stream box. The six sets of SNR measurement yield a median of $\sim 21$. We then add background stars to the simulations by estimating the stream density using a similar on-stream box and calculating the background density from the SNR.

Once we account for the background, for most simulated streams that have sub-streams, only the densest parts of the sub-streams can be distinguished from the background. For example, two segments of the sub-stream in figure \ref{fig:gd1comp1} can be identified, and they resemble the sub-structures found in GD-1. The example in \ref{fig:gd1comp1} is one of the better qualitative matches to the spur/blob features in GD-1 that we encountered in our simulations, but similar off-stream features are quite common in our simulation suite. Often they occur at somewhat larger separations from the main stream, where they would be more difficult to detect.

\begin{figure}
    \centering
    \includegraphics[width=\linewidth]{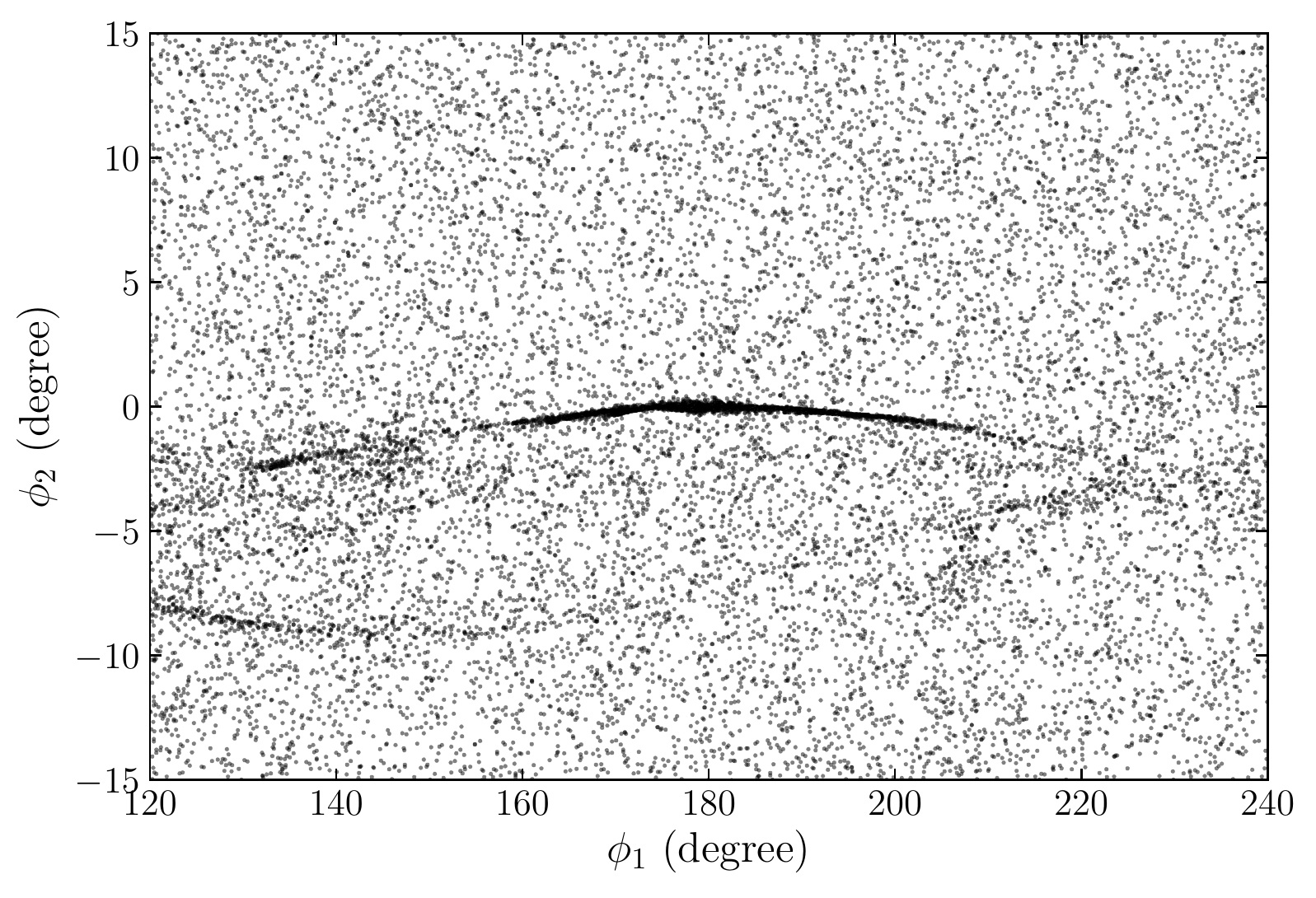}
    \caption{Example simulation where the accreted stream causes the appearance of a gap between $\phi_1 = 150^\circ$ and $160^\circ$ and a clump at $\phi_1 = 150^\circ$. The stream is generated from a globular cluster unbound from Sextans at $15.67 \, \mathrm{Gyr}$.}
    \label{fig:gd1comp2}
\end{figure}

\begin{figure}
    \centering
    \includegraphics[width=\linewidth]{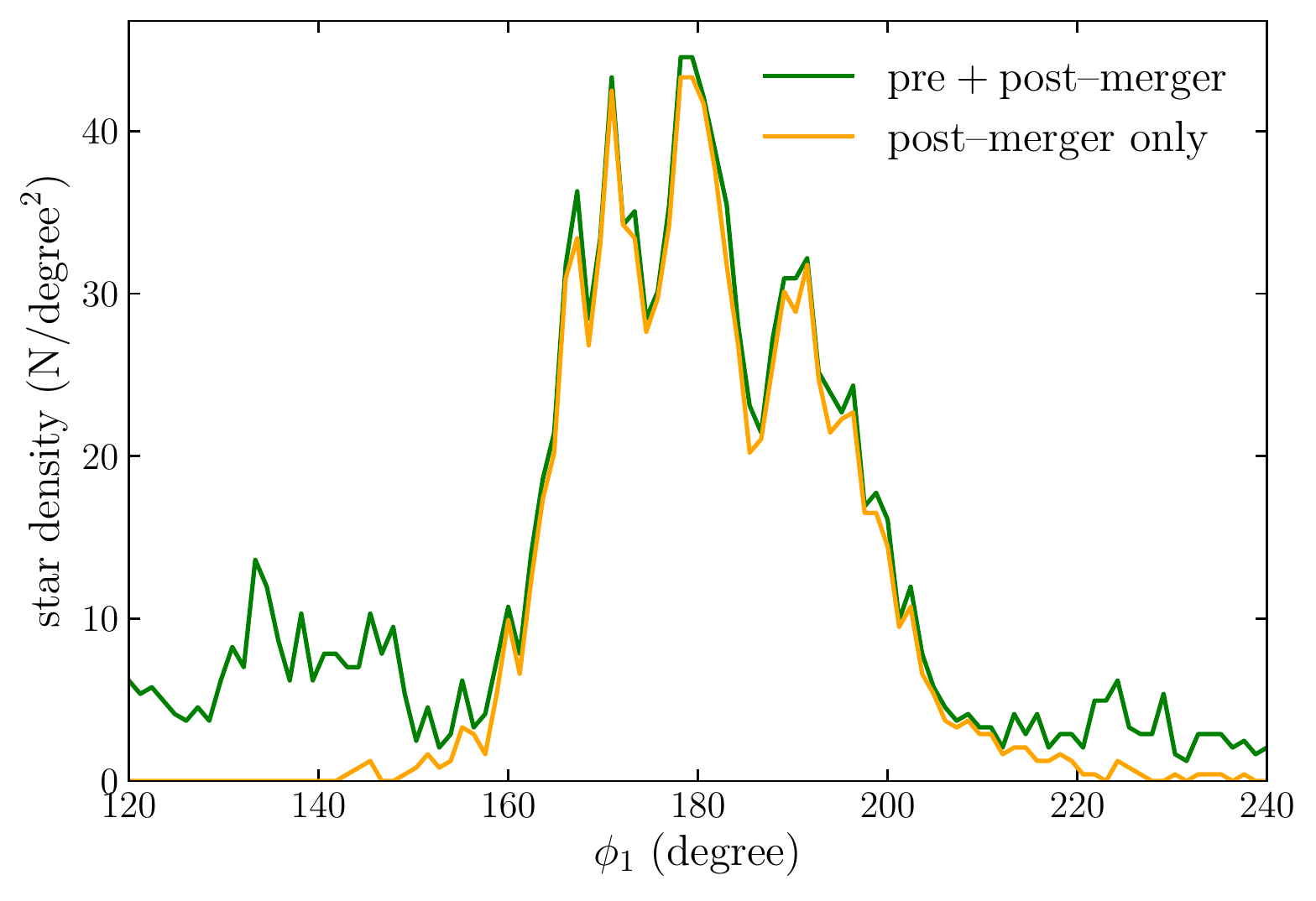}
    \caption{Area density of stars measured along the main track of the stream in figure \ref{fig:gd1comp2} (green line). A deficit of stars occurs around $\phi_1 = 155^\circ$. In contrast, the density of post-merger stars only (orange line) does not show a gap. Within the region where the post-merger, thin stream dominates the density, the effect of the accreted stream on the density is small.}
    \label{fig:main density}
\end{figure}

Another ubiquitous feature of detailed observations of stellar streams is one or more gaps. For example, GD-1 has several gaps \citep{price2018,deBoer2020}. Gaps are particularly exciting, because they are plausibly the result of encounters with dark matter subhalos \citep{Yoon2011,Carlberg2012,Bovy2017} and if interpreted as resulting from perturbations from baryonic and dark-matter substructures, the density of the GD-1 stream indeed provides evidence for the population of dark-matter subhalos down to $\approx 10^7\,M_\odot$ expected in the cold dark matter paradigm \citep{Banik2021}. However, it is possible that the interplay between the accreted and post-merger (thin) stream results in gaps that are not caused by any substructure in the mass distribution, but simply result from the stream's history. To investigate this, we look for gap features in our simulations along the track of the main stream. An example is shown in Figure \ref{fig:gd1comp2}: the simulated stream has a gap in the density along the main stream between $\phi_1 = 150^\circ$ and $\phi_1 = 160^\circ$. To obtain the density along the stream, we determine the main track as the median of the points in bins in $\phi_1$ and fit a fourth-order polynomial; a visual inspection of the result shows that this adequately locates the main track. Figure \ref{fig:main density} then displays density of stars along the stream as a function of $\phi_1$ as measured within one degree from the stream track. We show the density both for all stars and for just the stars in the post-merger stream to determine which parts of the density are due to the presence of the accreted stream. It is clear from Figure \ref{fig:main density} that the gap results from the accreted stream and that the accreted stream has little effect on the density in the area of high density dominated by the post-merger stream. Overall, the appearance of gaps such as that in Figure \ref{fig:main density} is relatively rare in our simulations and only a small fraction of our simulations display gaps that could be confused for dark-matter-subhalo gaps. This small occurence rate should be contrasted with the occurence rate of dark-matter-subhalo gaps, which is $\sim 1$ per stream \citep{Erkal2016}.

Thus, the ubiquitous appearance off-stream features in our simulations confirms the result of \citet{Malhan2019} that a ``cocoon'', or similar off-stream structure, is a hallmark of the accreted stellar stream. However, this structure is unlikely to affect the analyis of the density structure of the main stellar stream in terms of small-scale perturbations in the mass distribution, such as those coming from dark-matter subhalos.

\section{Conclusions}\label{sec:conclusions}

In this study, we simulate a number of accreted stellar streams with different initial and merger conditions. Globular-cluster stream progenitors are started off inside satellite galaxies where tidal stripping creates a pre-merger tidal stream. After the satellite merges with the Milky Way, this pre-merger stream becomes an accreted stream and post-merger tidal stripping of the globular cluster leads to a post-merger stream that is thin and simple as expected for a stream formed in a smooth, static potential. In the resulting configuration, the pre-merger stars in the accreted stream behave very differently from the post-merger stream stars. Accreted-stream members are spread over the entire sky and form 2D structures such as wider streams and sub-streams, and these structures become more complex as we allow the globular cluster to only be stripped from its parent satellite closer to the Milky Way center. The accreted stream that accompanies tidally-filling accreted globular clusters can therefore take on different forms, including but not limited to the ``cocoon'' feature first proposed by \citet{Malhan2019}.

The purpose of this paper is to study a wide range of accreted-stream behaviour and, as such, we have made various simplifications in the dynamics of the accreted streams to be able to consider many models in a reasonable amount of time. To obtain more quantitative results, a number of improvements can be adopted. First, we can initialize the parent orbits from a halo catalog of an $N$-body simulation, focusing on parent dwarfs that are fully disrupted at the present time, to avoid using surviving dwarf orbits, because surviving dwarf orbits may not be representative of the orbits of accreted Milky Way dwarf galaxies. Second, one can also properly simulate the $N$-body dynamics of a disrupting globular cluster in the host+dwarf potential, to generate the tidal stream consistently at all times, particularly during the tidal disruption of its parent galaxy. Moreover, since the whole process has a time span of $\sim 10$ Gyr, using a model for the time-evolving potential of the Milky Way instead of the static one will better mimic the evolution of streams after the merger. While these improvements would allow for more quantitative predictions of the structure of accreted streams to be made, our simulations suffice to understand the overall properties of accreted globular-cluster streams.

By creating mock observations of our simulations with similar characteristics as that of the observed GD-1 stream, which displays interesting off-track structure, we find that the off-track features around GD-1 can easily be explained as segments of the sub-streams that form for certain merger configurations in our simulations. In particular, our simulated streams often exhibit structures similar to the cocoon or spur/blob features identified around GD-1 and previously interpreted as caused by a perturbation from a globular cluster or compact dark-matter subhalo \citep{bonaca2020} or by the Sagittarius dwarf galaxy \citep{deBoer2020}.

Whether off-track features around observed streams can be uniquely attributed to the presence of an accreted-stream component remains unclear and requires further work to establish. The action distributions of accreted streams in Figures \ref{fig:largejr} and \ref{fig:largejp} suggest at least one way to distinguish between an accreted-stream nature versus a perturber-origin for off-track structures using the velocities of stars in the off-track features. A perturbation resulting in a spur or cocoon causes only a small change in the orbit of stars and the velocities of stars in the off-track features are therefore very similar to those of stars in the main stream. The pre-merger stars that make up the accreted stream, however, are influenced by the long-acting Milky-Way-–satellite system and thus have a much wider action and velocity distribution. \citet{bonaca2020} measure velocities of stars in the GD-1 spur and find that the radial velocities of these stars are consistent with those in the main stream. However, further work is required to determine whether this rules out the accreted-stream explanation. If the globular cluster has significant internal dynamical evolution resulting in internal mass segregation \citep[e.g.][]{Webb2016} then stars in the accreted stream that escaped long before the stars that make up the main stream may also have significantly different typical masses and other properties that allow their origin to be elucidated. This can be tested with deep observations of known streams by, e.g., the Rubin Observatory Legacy Survey of Space and Time (LSST) or the Nancy Grace Roman Space Telescope.

Further away from the post-merger stream, we can expect parts of accreted streams to show up as sub-streams that look like relatively narrow regular streams themselves. These can occur in parts of the sky and orbits that are far removed from the post-merger, thin stream, as shown in Figures \ref{fig:res basic} and \ref{fig:tshift}, and they would therefore not easily be associated with their post-merger counterpart. If the accreted stream is well populated through tidal stripping within its parent satellite, these sub-streams may have high surface densities and be detectable. Thus, this leads to the tantalizing possibility that some of the observed many thin streams without progenitors in the Milky Way are really part of a single or a few accreted streams, created in an ancient, massive merger. Recent spectroscopic observations have, for example, already shown that the disparate ATLAS and Aliqa Uma streams are in fact part of a single stream \citep{Li2021}. Because the stars in the accreted and post-merger stream originate in the same globular cluster, they should have very similar chemical abundances. Measurements of detailed chemical abundances along known streams, e.g. using future surveys such as MSE \citep{MSE}, will be able to determine whether seemingly disparate streams have the same progenitor through chemical tagging.

\section*{Acknowledgements}

It is a pleasure to thank Ray Carlberg and an anonymous referee for helpful comments on this paper. YQ, YA, and JB acknowledge financial support from NSERC (funding reference number RGPIN-2020-04712) and an Ontario Early Researcher Award (ER16-12-061). This work has made use of data from the European Space Agency (ESA) mission
{\it Gaia} (\url{https://www.cosmos.esa.int/gaia}), processed by the {\it Gaia}
Data Processing and Analysis Consortium (DPAC,
\url{https://www.cosmos.esa.int/web/gaia/dpac/consortium}). Funding for the DPAC
has been provided by national institutions, in particular the institutions
participating in the {\it Gaia} Multilateral Agreement.

\section*{Data Availability}

This research made use of data on the present phase-space positions of satellite galaxies from \citet{Fritz2018} and \citet{Helmi18} determined from \citet{gaia} data.

\bibliographystyle{mnras}
\bibliography{streams}

\bsp	
\label{lastpage}
\end{document}